\documentclass[12pt,a4paper]{article}
\usepackage[left=2.5cm,right=2.5cm,top=2.5cm,bottom=2.5cm]{geometry}
\usepackage{comment}
\usepackage[ruled,vlined]{algorithm2e}
\usepackage{amsfonts}
\usepackage{amsmath}
\usepackage{amssymb}
\usepackage{amsthm}
\usepackage{array}
\usepackage{authblk}
\usepackage{bm}
\usepackage{booktabs}
\usepackage{caption}
\usepackage{color}
\usepackage{graphicx}
\usepackage[colorlinks,linkcolor=blue,citecolor=blue,urlcolor=blue]{hyperref} 
\usepackage{multirow}
\usepackage[authoryear,semicolon]{natbib}
\usepackage{physics}
\usepackage{siunitx}
\usepackage{standalone}
\usepackage{setspace}
\usepackage{subfigure}
\usepackage{tikz}
\usepackage[normalem]{ulem}
\usepackage{url}
\usepackage{lineno}
\allowdisplaybreaks
\usepackage{placeins} 

\newcommand*\patchAmsMathEnvironmentForLineno[1]{%
  \expandafter\let\csname old#1\expandafter\endcsname\csname #1\endcsname
  \expandafter\let\csname oldend#1\expandafter\endcsname\csname end#1\endcsname
  \renewenvironment{#1}%
     {\linenomath\csname old#1\endcsname}%
     {\csname oldend#1\endcsname\endlinenomath}}%
\newcommand*\patchBothAmsMathEnvironmentsForLineno[1]{%
  \patchAmsMathEnvironmentForLineno{#1}%
  \patchAmsMathEnvironmentForLineno{#1*}}%

\AtBeginDocument{%
\patchBothAmsMathEnvironmentsForLineno{equation}%
\patchBothAmsMathEnvironmentsForLineno{align}
\patchBothAmsMathEnvironmentsForLineno{flalign}%
\patchBothAmsMathEnvironmentsForLineno{alignat}%
\patchBothAmsMathEnvironmentsForLineno{gather}%
\patchBothAmsMathEnvironmentsForLineno{multline}%
}

\newcolumntype{x}[1]{>{\centering\arraybackslash\hspace{0pt}}p{#1}}
\newcommand{\E}{\text{E}}

\newcommand{\Var}{\text{Var}}
\newcommand{\bdiag}{\text{bdiag}}

\newcommand{\txd}{\text{d}}

\newcommand{\N}{\mathcal{N}}

\newcommand{\trc}{\text{tr}}
\newcommand{\const}{\text{const}}

\newcommand{\betavec}{\boldsymbol{\beta}}
\newcommand{\epsilonvec}{\boldsymbol{\epsilon}}

\newcommand{\muvec}{\boldsymbol{\mu}}

\newcommand{\sigmavec}{\boldsymbol{\sigma}}
\newcommand{\thetavec}{\boldsymbol{\theta}}

\newcommand{\zerovec}{\boldsymbol{0}}

\newcommand{\Sigmavec}{\boldsymbol{\Sigma}}

\newcommand{\bvec}{\mathbf{b}}

\newcommand{\mvec}{\mathbf{m}}

\newcommand{\uvec}{\mathbf{u}}
\newcommand{\vvec}{\mathbf{v}}
\newcommand{\wvec}{\mathbf{w}}
\newcommand{\xvec}{\mathbf{x}}
\newcommand{\yvec}{\mathbf{y}}
\newcommand{\zvec}{\mathbf{z}}

\newcommand{\Cvec}{\mathbf{C}}

\newcommand{\Ivec}{\mathbf{I}}
\newcommand{\Lvec}{\mathbf{L}}
\newcommand{\Mvec}{\mathbf{M}}

\newcommand{\Svec}{\mathbf{S}}
\newcommand{\Tvec}{\mathbf{T}}
\newcommand{\Uvec}{\mathbf{U}}
\newcommand{\Vvec}{\mathbf{V}}
\newcommand{\Xvec}{\mathbf{X}}

\newcommand{\Zvec}{\mathbf{Z}}

\newtheorem{thm}{Theorem}

\begin{document}

\title{\textbf{\large{Random Effects Misspecification and its Consequences for Prediction in Generalized Linear Mixed Models}}}  
  \author[1,*]{\normalsize{Quan Vu}}
  \author[1]{\normalsize{Francis K. C. Hui}}
  \author[2]{\normalsize{Samuel Muller}}
  \author[1]{\normalsize{A. H. Welsh}}
  \affil[1]{\small{Research School of Finance, Actuarial Studies and Statistics, The Australian National University, Australia}}
  \affil[2]{\small{School of Mathematical and Physical Sciences, Macquarie University, Australia}}
  \affil[*]{\small{Corresponding author: quan.vu@anu.edu.au}}
  \date{}
  \maketitle

\begin{abstract}
When fitting generalized linear mixed models (GLMMs), one important decision to make relates to the choice of the random effects distribution. As the random effects are unobserved, misspecification of this distribution is a real possibility. In this article, we investigate the consequences of random effects misspecification for point prediction and prediction inference in GLMMs, a topic on which there is considerably less research compared to consequences for parameter estimation and inference. We use theory, simulation, and a real application to explore the effect of using the common normality assumption for the random effects distribution when the correct specification is a mixture of normal distributions, focusing on the impacts on point prediction,  mean squared prediction errors (MSEPs), and prediction intervals. We found that the optimal shrinkage is different under the two random effect distributions, so is impacted by misspecification. The unconditional MSEPs for the random effects are almost always larger under the misspecified normal random effects distribution, especially when cluster sizes are small. Results for the MSEPs conditional on the random effects are more complicated, but they remain generally larger under the misspecified distribution when the true random effect is close to the mean of one of the component distributions in the true mixture distribution. Results for prediction intervals indicate that overall coverage probability is not greatly impacted by misspecification.
\\

\noindent
{\it Keywords:} Clustered data, Empirical best predictor, Longitudinal data, Mean squared error of prediction, Prediction inference.

\end{abstract}


\section{Introduction} \label{sec:intro}

Generalized linear mixed models \citep[GLMMs,][]{jiang2007linear,mcculloch2011generalized} are widely used in many disciplines to model correlations between observations. 
The correlations are induced through unobserved random effects, and as such, one important decision to make when fitting GLMMs is what distribution to assume for the random effects. As a motivating example, we consider a study to examine the effect of time spent in the workforce on hourly wages, where we found evidence the random effects distribution exhibits deviations from the usually assumed normal distribution; see Section \ref{sec:case_study} for more details. Such examples raise the question of what the consequences are for estimation, inference, and prediction when we misspecify the random effects distribution in GLMMs.

The vast majority of applications and statistical software packages for fitting GLMMs, such as \texttt{lme4} \citep{bates2015} and \texttt{glmmTMB} \citep{brooks2017}, assume a normal distribution for the random effects. Yet the issue of how random effects misspecification 
impacts on estimation and inference in GLMMs remains unresolved despite considerable research in the area.
\cite{richardson1994asymptotic} and  \cite{verbeke1997effect} found that assuming normality in linear mixed models (LMMs),
the maximum likelihood estimators of both fixed effects and variance components are consistent for the corresponding true parameter values, even when the underlying random effects distribution is not normal.
In finite samples though, \cite{hui2021random} demonstrated substantial bias and mean squared error in estimators of the variance components under misspecification of the random effects distribution in LMMs. 
Turning to GLMMs, \cite{mcculloch2011misspecifying} and \cite{neuhaus2013estimation} showed that estimators of both fixed effects and variance components are largely robust to random effects misspecification. This contrasts with \cite{heagerty2001misspecified} and \cite{litiere2008impact}, who found the bias in point estimators of all model parameters in GLMMs can be substantial when the true random effects distribution is severely non-normal. 

As noted by \citet{hui2021random}, one particular area where the impact of random effects misspecification in GLMMs remains understudied is point prediction and prediction inference. 
There are several approaches to obtain the predicted random effects, and in this paper we focus on the popular best prediction approach, which minimizes mean squared prediction error.
Not surprisingly, the comparably little research that has been conducted also presents conflicting results. 
\cite{verbeke1996linear} and \cite{mcculloch2011misspecifying} showed empirically that for independent-cluster GLMMs, the normality assumption results in too normal-like predicted random effects distribution when the true distribution is not normal. They concluded that determining the true shape of the random effects distribution from predictors of the random effects is a challenging task. 
This contrasts with results of \cite{hui2021random}, who found for LMMs that the shape of the predicted random effects distribution can exhibit clear non-normality, and resembles the true shape when the cluster size is sufficiently large; see also the recent related asymptotic results by \cite{lyu2022asymptotics} and \cite{ning2024asymptotic} which support this idea.
For the predictors themselves, \cite{agresti2004examples} found substantial bias in the best predictors when the true random effects distribution was a mixture with a large variance. This contrasts with \cite{mcculloch2011prediction}, who found that for both best prediction and prediction using the mode, the unconditional mean square error of prediction (UMSEP) was fairly robust to the random effects distribution, provided the true random effects variance was not too large and cluster sizes were moderate.
The UMSEP is widely used for uncertainty quantification and constructing prediction intervals, particularly in small-area estimation \citep[e.g.,][]{rao2015small}. However, other researchers have proposed using the conditional mean squared error of prediction (CMSEP) for cluster-level prediction, either conditioning on the random effects \citep[e.g.,][]{zheng2023frequentist} or conditioning on the responses from that cluster \citep{booth1998standard,lee2011prediction,korhonen2023fast}. For instance, in our motivating example of hourly wages data, it is of interest to predict conditioned on high earning individuals.
How random effects misspecification impacts these different forms of the MSEP, and thus different types of prediction intervals for GLMMs, is largely unknown.

In this article, we study the consequences of random effects misspecification 
for point prediction and prediction inference in independent-cluster GLMMs. 
We consider its impact on both the UMSEP and the less studied CMSEP, comparing prediction performance 
when the true random effects distribution is normal with that of a finite mixture of normal distributions. Random effects misspecification here thus refers to assuming a normal distribution when the true distribution is a mixture of normals. 
While other non-normal random effects could be considered \citep[e.g., see the works of][]{litiere2008impact,mcculloch2011misspecifying,hui2021random}, we purposely use a finite mixture of normals since: 1) it can represent a wide range of continuous distributions including those with multimodality, skewness, and heavy-tailed behavior \citep[e.g.,][]{nguyen2019approximations}; see also \cite{komarek2008generalized, liu2008likelihood, pan2020generalized} where finite mixture random effects distributions have been used in GLMMs; and 2) 
unlike most other choices, a finite mixture of normal random effects distributions allows us to produce closed-forms for the CMSEP and other quantities under the true random effects distribution, thereby offering opportunities for analytical insights into the impacts of  misspecification.
We also consider the impact on prediction intervals constructed from the point predictors and their corresponding unconditional and conditional MSEPs. 

We summarize our results into three main findings. First, the UMSEP 
obtained under the misspecified normal distribution is consistently larger, with the differences being more pronounced when cluster sizes are small and the number of clusters is large. Second, the CMSEP is also larger under the misspecified normal random effects distribution, although the extent depends on the true finite mixture distribution and the response distribution. For instance, if the true random effects distribution is asymmetric, then CMSEPs under misspecification are larger when the true random effect is near the mean of one of the mixture components. Third, prediction intervals based on MSEPs are generally wider when using the misspecified normal distribution. While coverage probabilities for prediction intervals using UMSEPs vary widely across clusters (which is not too surprising given UMSEP is based on averaging over all clusters) and regardless of the true random effects distribution, prediction intervals using CMSEPs tend to achieve nominal coverage even under misspecification.

Overall, our results suggest that random effects misspecification can have consequences for point prediction and prediction inference in GLMMs, and we encourage greater caution when it comes to adopting the standard normality assumption in GLMMs. If fitting a mixed model with this assumption leads to clear non-normal random effect predictions, then the assumption should be reconsidered and modified appropriately. Additionally, if the dataset has mostly small clusters (with ``small'' depending on the response distribution), the impact of misspecification will be more severe.

The remainder of this article is organized as follows: 
Section \ref{sec:pred} reviews independent-cluster GLMMs and the (empirical) best predictors of the random effects, before presenting forms for the UMSEP and CMSEP. Section \ref{sec:pred_lmm} focuses on the special case of LMMs and derives closed-form expressions for the predictors and two MSEPs under the normal and mixture of normal distributions for the random effects. These derivations are supported by simulation studies with binary and count responses in Section \ref{sec:sim}, and an application to the motivating longitudinal hourly wages data in Section \ref{sec:case_study}. Section \ref{sec:conclusion} offers some concluding remarks.

\section{Independent-cluster GLMMs and prediction}\label{sec:pred}

Consider a set of independent clusters $i = 1, \dots, m$, where in cluster $i$ we observe the response vector $\yvec_i = (y_{i1}, \dots, y_{i n_i})^\top$, a vector of fixed effect covariates $\xvec_{ij} = (x_{ij1},\ldots, x_{ijq_f})^\top$, and a set of random effect covariates $\zvec_{ij} = (z_{ij1},\ldots,z_{ijq_r})^\top$ for observation $j=1,\ldots,n_i$. We will assume the first elements $x_{ij1} = z_{ij1} = 1$ to represent a fixed and random intercept term, respectively. 
We fit an independent-cluster GLMM to this data, in which the conditional distribution of $y_{ij}$ given a vector of random effects $\uvec_i = (u_{i1},\ldots,u_{iq_r})^\top$ comes from the exponential family of distributions. That is, $p(y_{ij}\mid\betavec, \phi, \uvec_i) = \exp[ \{y_{ij}\vartheta_{ij} - a(\vartheta_{ij}) \} /\phi + b(y_{ij},\phi)]$ for known functions $a(\cdot)$ and $b(\cdot)$, and a dispersion parameter $\phi > 0$ that may or may not require estimation depending on the precise distribution of interest. 
The mean of the distribution, denoted here as $\mu_{ij} = a'(\vartheta_{ij})$, 
is modeled as a linear combination of the fixed and random effect covariates through a link function $g(\cdot)$ i.e.,
$
g(\mu_{ij}) = \eta_{ij} = \xvec_{ij}^\top \betavec + \zvec_{ij}^\top \uvec_i,
$
where $\betavec$ denotes the $q_f$-vector of fixed effect coefficients. Finally, for $i = 1,\ldots,m$, the random effects $\uvec_i$ are assumed to follow a distribution $p(\uvec_i \mid \sigmavec)$ with parameter vector $\sigmavec$, and which satisfies $\E(\uvec_i) = \bm{0}_i$; the zero expectation is required for parameter identifiability. We assume $\text{Cov}(\uvec_i) = \Sigmavec$, which is parametrized by $\sigmavec$. 
As reviewed in Section \ref{sec:intro}, a common choice implemented in a variety of software for fitting GLMMs assumes $p(\uvec_i \mid \sigmavec)$ follows a $q_r$-dimensional normal distribution, $p(\uvec_i \mid \sigmavec) = \mathcal{N}_{q_r}(\bm{0}_{q_r}, \bm{\Sigma})$ where $\sigmavec = \text{chol}(\bm{\Sigma})$ in the case of an unstructured covariance matrix. Throughout this article, we define random effects misspecification as occurring when the true form of $p(\uvec_i \mid \sigmavec)$ does not follow such a normal distribution (although it has the same first two moments), but we assume normality when fitting the GLMM.
Assuming the responses $y_{ij}$ are conditionally independent given $\uvec_i$, the marginal log-likelihood of the independent-cluster GLMM is given by
$
\log L(\thetavec) = \sum_{i=1}^m \log \left\{\int \prod_{j=1}^{n_i} p(y_{ij}\mid\betavec, \phi, \uvec_i) p(\uvec_i \mid \sigmavec) \txd \uvec_i \right\},
$
where $\thetavec = (\betavec^\top, \phi, \sigmavec^\top)^\top$ denotes the vector of model parameters. The integral in the marginal log-likelihood function generally does not possess a tractable form, and for likelihood-based estimation there are a number of ways by which $\log L(\thetavec)$ can be maximized, most of which involve some approximation to the integral \citep[see for instance][among many others]{breslow1993approximate,wolfinger1993laplace,ormerod2012gaussian,brooks2017}.

\subsection{Random effect predictions} \label{subsec:randomeffectspredictions}
For point prediction of the random effects $\uvec_i$, arguably the most popular approach which we investigate below is the so-called best predictor \citep[BP, e.g.][]{booth1998standard, jiang2003empirical}, defined as the conditional expectation of the random effects given the observed data, 
\begin{align}\label{eq:BP}
\wvec_i(\thetavec) &= \E[\uvec_i \mid \yvec_i] = \int \uvec_i p(\uvec_i \mid \yvec_i, \thetavec) \txd\uvec_i = \frac{\int \uvec_i \prod_{j=1}^{n_i} p(y_{ij}\mid\betavec, \phi, \uvec_i) p(\uvec_i \mid \sigmavec) \txd\uvec_i}{\int  \prod_{j=1}^{n_i} p(y_{ij}\mid\betavec, \phi, \uvec_i) p(\uvec_i \mid \sigmavec) \txd\uvec_i}.
\end{align}

To understand how random effects misspecification impacts the BP, we consider two possible cases as follows: First, let $p_0(\uvec_i \mid \sigmavec)$ denote the true density of the random effects, which throughout the article we set to be a finite mixture of normal distributions. That is, $p_0(\uvec_i\mid \sigmavec) = \sum_{k=1}^c \pi_k p_k(\uvec_i | \muvec_{k},\sigmavec_{k}) =  \sum_{k=1}^c \pi_k \mathcal{N}_d(\uvec_i | \muvec_{k}, \Sigmavec_{k})$, where $\muvec_{k}$ and $\Sigmavec_{k}$ denote the mean vector and covariance matrix respectively for the $k$th mixture component, $\bm{\pi} = (\pi_1,\ldots,\pi_c)^\top$ denotes the vector of mixture probabilities, $c$ denotes the number of mixture components, and $\sigmavec = (\bm{\pi}^\top, \muvec_{1}^\top,\ldots,\muvec_{c}^\top,\sigmavec_{1}^\top,\ldots,\sigmavec_{c}^\top)^\top$
Note it is necessary to constrain $\sum_{k=1}^c \pi_k \muvec_{k} = \zerovec_d$ and $\sum_{k=1}^c \pi_k (\Sigmavec_{k} + \muvec_{k} \muvec^\top_{k}) = \Sigmavec$ to ensure the true random effects distribution continues to have zero expectation and covariance matrix $\bm{\Sigma}$. Let $\thetavec_0 = (\betavec_0^\top, \phi_0, \sigmavec_0^\top)^\top$ denote the true parameters in the GLMM under this true mixture distribution. It follows that the conditional distribution of the random effects given the observed data is given by
\begin{align} 
p_{0}(\uvec_i \mid \yvec_i, \thetavec_0) &= \frac{\prod_{j=1}^{n_i} p(y_{ij}\mid\betavec_0, \phi_0, \uvec_i) \sum_{k=1}^c \pi_{0k} p_k(\uvec_i\mid\muvec_{0k},\sigmavec_{0k}) }{\int  \prod_{j=1}^{n_i} p(y_{ij}\mid\betavec_0, \phi_0, \uvec_i) \sum_{k=1}^c \pi_{0k} p_k(\uvec_i\mid\muvec_{0k},\sigmavec_{0k}) \txd\uvec_i} \nonumber  \\
&= \frac{\sum_{k=1}^c \pi_{0k} p_k(\yvec_i \mid \thetavec_0) p_k(\uvec_i \mid \yvec_i, \thetavec_0)}{\sum_{k=1}^c \pi_{0k} p_k(\yvec_i \mid \thetavec_0)} \label{eq:BP_true},
\end{align}
where $p_k(\yvec_i \mid \thetavec_0) = \int \prod_{j=1}^{n_i} p(y_{ij}\mid\betavec_0, \phi_0, \uvec_i) p_k(\uvec_i\mid\muvec_{0k},\sigmavec_{0k}) \txd\uvec_i$ can be regarded as the marginal likelihood for the $i$th cluster assuming it belongs to mixture component $k = 1,\ldots,c$, and $p_k(\uvec_i \mid \mathbf{y}_i, \thetavec_0) = \prod_{j=1}^{n_i} p(y_{ij}\mid\betavec_0, \phi_0, \uvec_i) p_k(\uvec_i \mid \muvec_{0k}, \sigmavec_{0k}) / p_k(\yvec_i\mid \thetavec_0)$ is the corresponding conditional distribution given the observed data.
Based on \eqref{eq:BP_true}, we deduce that the true BP $\wvec_{0i}(\thetavec_0) = \int \uvec_i p_0(\uvec_i \mid \yvec_i, \thetavec_0) \txd\uvec_i$ can be written as a weighted sum of best predictors under each mixture component distribution, where the weights are proportional to $\pi_{0k} p_k(\yvec_i \mid \thetavec_0)$. 

Next, let $p_{*}(\uvec_i \mid \sigmavec)$ denote the misspecified normal random effects distribution, $p_{*}(\uvec_i \mid \sigmavec) = \mathcal{N}(0, \Sigmavec)$, and let $\thetavec_{*} = (\betavec_{*}^\top, \phi_{*}, \sigmavec_{*}^\top)^\top$ denote the corresponding vector of pseudo-true model parameters for this misspecified GLMM. To be clear, the pseudo-true parameters $\thetavec_{*}$ are defined as the set of parameters which minimize the Kullback-Leibler divergence between the marginal likelihood of the true and misspecified GLMMs; see \citet{white1982maximum} for a general definition of pseudo-true parameters and \citet{verbeke1997effect, heagerty2001misspecified, hui2022assuming} for discussion specifically relating to (G)LMMs. It follows that
\begin{equation*}
p_{*}(\uvec_i \mid \yvec_i, \thetavec_{*}) = \frac{ \prod_{j=1}^{n_i} p(y_{ij}\mid\betavec_{*}, \phi_{*}, \uvec_i) p_{*}(\uvec_i \mid \sigmavec_{*}) } { \int \prod_{j=1}^{n_i} p(y_{ij}\mid\betavec_{*}, \phi_{*}, \uvec_i) p_{*}(\uvec_i \mid \sigmavec_{*}) \txd\uvec_i }, 
\end{equation*}
where the denominator is the marginal likelihood corresponding to the $i$th cluster for the misspecified GLMM. The resulting misspecified BP $\wvec_{*i}(\thetavec_{*}) = \int \uvec_i p_{*}(\uvec_i \mid \yvec_i, \thetavec_{*}) \txd\uvec_i$ thus takes the generic form given in \eqref{eq:BP}. 

Note the true and misspecified best predictors are functions of the true and pseudo-true parameters, respectively. Upon replacing these with corresponding likelihood-based estimates based on fitting the true and misspecified GLMMs, which we denote here as $\hat{\thetavec}_0$ and $\hat{\thetavec}_*$ respectively, we obtain estimates often referred to as empirical best predictors \citep[EBPs,][]{booth1998standard, jiang2003empirical}. For the remainder of this article, we shall write the generic form of EBPs based on \eqref{eq:BP} as $\hat \wvec_i = \wvec_i(\hat{\thetavec})$ given estimates $\hat{\thetavec}$, the true EBP i.e., the EBP for the GLMM with the true random effects distribution, as $\hat \wvec_{0i} = \wvec_{0i}(\hat{\thetavec}_0)$, and the misspecified EBP i.e., the EBP for the GLMM with the misspecified distribution, as $\hat \wvec_{*i} = \wvec_{*i}(\hat{\thetavec}_{*})$.

\subsection{Mean squared prediction errors}\label{sec:msep}
Measuring the uncertainty of the EBP is important for prediction inference, and in this article we accomplish this through the mean squared error of prediction \citep[MSEP; for other examples where this is used, see][]{das2004mean, cantoni2017random, flores2019bootstrap}, noting this is often utilized to construct prediction intervals for the random effects.
This section introduces three different flavors of the MSEP for the random effects $\uvec_i$.

\underline{Unconditional mean squared error of prediction:} The UMSEP of $\uvec_i$, which is evaluated over the marginal distribution of $\yvec$,
is given by the diagonal entries of the matrix
\begin{equation}\label{eq:umsep}
\text{UMSEP} = \E[(\hat \wvec_i - \uvec_i)(\hat \wvec_i - \uvec_i)^\top] = \Uvec_1 + \Uvec_2 + \Uvec_3,
\end{equation}
where $\Uvec_1 = \E[(\wvec_i - \uvec_i)(\wvec_i - \uvec_i)^\top]$, $\Uvec_2 = \E[(\hat \wvec_i - \wvec_i)(\hat \wvec_i - \wvec_i)^\top]$, and $\Uvec_3 = \E[(\hat \wvec_i - \wvec_i)(\wvec_i - \uvec_i)^\top + (\wvec_i - \uvec_i)(\hat \wvec_i - \wvec_i)^\top]$. The term $\Uvec_1$ represents the error arising from using the BP, and its diagonal entries are the MSEPs when we know the true/pseudo-true parameters. The term $\Uvec_1$ also depends on the cluster size $n_i$, and tends to zero when $n_i \rightarrow \infty$.
The term $\Uvec_2$ represents the error arising from estimating the parameters, and converges to zero when the number of clusters $m \rightarrow \infty$.  Finally, the term $\Uvec_3$ involves the interaction between the prediction and the estimation. We refer to \citet{zheng2021frequentist} and \citet{ning2024asymptotic} for more details on the asymptotic behavior of these terms in GLMMs.

In the setting of longitudinal studies when the number of clusters $m$ is large relative to the cluster sizes $n_i$, we can assume $\Uvec_2$ tends to zero while $\Uvec_1$ remains non-trivial. Importantly, the term $\Uvec_1$ is most sensitive to misspecification of the random effects distribution, as $\wvec_i$ is dependent on the choice of $p(\uvec_i \mid \sigmavec)$; see Sections \ref{sec:pred_lmm} and \ref{sec:sim} for evidence of this. On the other hand, when $m$ is small relative to $n_i$, then terms $\Uvec_2$ and $\Uvec_3$ can still be non-trivial in finite samples; this will be shown in the simulation studies in Section \ref{sec:sim}. 

For the special case of LMMs, the UMSEP has been studied under the normal random effects assumption \citep[e.g.,][]{kackar1984approximations, prasad1990estimation}. In that case and when the random effects distribution is truly normally distributed, the term $\Uvec_1$ has a closed-form solution. In the following section, we show that even when the true random effects distribution is a mixture of normal distributions, $\Uvec_1$ still has a closed-form solution.

\underline{Conditional mean squared error of prediction:} The CMSEP is defined here as the MSEP given the random effects for cluster $i$. This quantity may be of interest when we are interested in prediction for a particular subset of the clusters or prediction for the mean or the sum of responses for a specific cluster \citep[e.g.,][]{rao2015small}. 
For cluster $i$, the CMSEP of the random effects $\uvec_i$  is given by the diagonal entries of the following matrix
\begin{equation}\label{eq:cmsep}
\text{CMSEP} = \E[(\hat \wvec_i - \uvec_i)(\hat \wvec_i - \uvec_i)^\top \mid \uvec_i] = \Cvec_1 + \Cvec_2 + \Cvec_3,
\end{equation}
where $\Cvec_1 = \E[(\wvec_i - \uvec_i)(\wvec_i - \uvec_i)^\top \mid \uvec_i]$, $\Cvec_2 = \E[(\hat \wvec_i - \wvec_i)(\hat \wvec_i - \wvec_i)^\top \mid \uvec_i]$, and $\Cvec_3 = \E[(\hat \wvec_i - \wvec_i)(\wvec_i - \uvec_i)^\top + (\wvec_i - \uvec_i)(\hat \wvec_i - \wvec_i)^\top \mid \uvec_i]$. 
Each term $\Cvec_1, \Cvec_2, \Cvec_3$ in \eqref{eq:cmsep} is analogous to the terms $\Uvec_1, \Uvec_2, \Uvec_3$ in \eqref{eq:umsep}: the term $\Cvec_1$ is dependent on $n_i$ and converges to zero as the cluster size grows, while the term $\Cvec_2$ converges to zero as the number of clusters $m$ grows. 
Similar to the term $\Uvec_1$ in the UMSEP, the term $\Cvec_1$ in the CMSEP is most affected by misspecification, with the impact more evident when the cluster size is smaller.
For LMMs, we can again derive a closed-form expression for $\Cvec_1$ under both the misspecified normal and the mixture random effects distributions, thus enabling some analytical insight as we shall see in the following section.

\underline{Bootstrap estimates of the MSEPs:}
For a general situation where the response distribution is non-normal, there is usually no closed-form solution for the expressions in \eqref{eq:umsep} and \eqref{eq:cmsep}.
Even in LMMs, the term $\Uvec_2$ (and similarly $\Cvec_2$) is not available in closed form and usually has to be approximated using a Taylor expansion \citep[e.g.,][]{kackar1984approximations}.
With this in mind, in this paper we also consider a more general, practical approach to estimate the UMSEP and CMSEP via the parametric bootstrap \citep{chatterjee2008parametric, flores2019bootstrap}. 
A full parametric bootstrap algorithm to estimate the UMSEP and CMSEP is provided in the Supplementary Material. Briefly, it involves either resampling the random effects from the estimated random effects distribution (for UMSEP) or conditioning on  the set of predicted random effects (for CMSEP).

\section{Random effects misspecification in LMMs}\label{sec:pred_lmm}

In this section, we examine the special case of independent-cluster linear mixed models, as this is one of the primary cases for which we can arrive at a closed-form expression and hence analytical insights for $\Uvec_1$ and $\Cvec_1$ in \eqref{eq:umsep} and \eqref{eq:cmsep}, respectively. Recall from the preceding section that these terms are the most sensitive, in their respective MSEPs, to misspecification of the random effects distribution.

Consider the independent cluster LMM 
$y_{ij} = \xvec_{ij}^\top \betavec + \zvec_{ij}^\top \uvec_i + \epsilon_{ij}, $
where 
$\epsilon_{ij}$ is assumed to be a normally distributed error with mean zero and variance $\tau^2$. Throughout the developments below, we will assume the distribution of $\epsilon_{ij}$ is correctly specified. 
Next, suppose the true distribution of the random effects is a mixture of normal distributions, such that analogous to below \eqref{eq:BP} we can write  
$p_0(\uvec_i | \sigmavec_0) = \sum_{k=1}^c \pi_{0k} \N(\uvec_i | \Lvec_0 \muvec_{0k}, \Lvec_0 \Sigmavec_{0k} \Lvec_0^\top)$, with constraints $\sum_{k=1}^c \pi_{0k} \muvec_{0k} = \zerovec $ and $\sum_{k=1}^c \pi_{0k} (\muvec_{0k} \muvec^\top_{0k} + \Sigmavec_{0k}) = \Ivec $. The corresponding misspecified random effects distribution is denoted by $p_{*}(\uvec_i | \sigmavec_*) = \N(0, \Sigmavec_{*})$.  We define $\Lvec = \text{chol}(\Sigmavec)$ and $\Lvec_{*} = \text{chol}(\Sigmavec_{*})$, and 
begin by establishing the following result on the equality between the pseudo-true and true parameters for LMMs. The proof is given in Supplementary Material S1.

\begin{thm} \label{thm:LMMequivalence}
If the true random effects distribution follows a (standardized) mixture of normal distributions in an independent cluster LMM, then the pseudo-true model parameter $\thetavec_{*} = (\betavec^\top_{*}, \Lvec^\top_{*}, \tau_{*}^2)^\top$ from a LMM assuming a misspecified normal random effects distribution is equal to the true parameter  $\thetavec_{0} = (\betavec^\top_{0}, \Lvec^\top_{0}, \tau_{0}^2)^\top$ . 
\end{thm}

In the next two sections, we use this result to establish formulas for the BPs and their corresponding MSEPs. In particular, we will see that even though the pseudo-true and true parameters are equal in the LMM setting, the corresponding BPs and the MSEPs can be quite different, and prediction can be greatly impacted by random effects misspecification even when estimation is not. In what follows, we use $\thetavec = (\betavec^\top, \Lvec^\top, \tau^2)^\top$ to denote both $\thetavec_{*}$ and $\thetavec_0$ in light of Theorem \ref{thm:LMMequivalence}.

\subsection{Best predictors}\label{sec:lmm_predictor}
Applying the results in Section \ref{subsec:randomeffectspredictions} to the special case of the best predictor $\wvec_{*i}$ of the random effect vector $\uvec_i$ under the misspecified LMM (also known as the best unbiased linear predictor or BLUP in this case), we can show that
\begin{align}\label{eq:BLUP_mis}
\begin{split}
\wvec_{*i}(\thetavec) = \E_{*}(\uvec_i \mid \yvec_i) &= \{(\Lvec \Lvec^\top)^{-1} + \tau^{-2} \Zvec^\top_i \Zvec_i\}^{-1} \tau^{-2} \Zvec^\top_i (\yvec_i - \Xvec_i \betavec) \\
&= \{(\Lvec \Lvec^\top)^{-1} + \tau^{-2} \Zvec^\top_i \Zvec_i\}^{-1} \tau^{-2} \Zvec^\top_i (\Zvec_i \uvec_i + \epsilonvec_i),
\end{split}
\end{align}
where $\Xvec_i = (\xvec_{i1}, \dots, \xvec_{i n_i})^\top$, and $\Zvec_i = (\zvec_{i1}, \dots, \zvec_{i n_i})^\top$.
On the other hand, the best predictor $\wvec_{0i}$ under the true LMM is given by
\begin{equation}\label{eq:BLUP_true}
\wvec_{0i}(\thetavec) = \frac{\sum_{k=1}^c \pi_{0k} p_k(\yvec_i) \mvec_k}{\sum_{k=1}^c \pi_{0k} p_k(\yvec_i)},
\end{equation}
where $\mvec_k = \{ (\Lvec \Sigmavec_{0k} \Lvec^\top )^{-1} + \tau^{-2} \Zvec^\top_i \Zvec_i \}^{-1} \{(\Lvec \Sigmavec_{0k} \Lvec^\top)^{-1} \Lvec \muvec_{0k} + \tau^{-2} \Zvec^\top_i (\yvec_i - \Xvec_i \betavec)\},$
and $p_k(\yvec_i) = \N(\Xvec_i \betavec + \Zvec_i \Lvec \muvec_{0k}, \tau^2 \Ivec + \Zvec_i (\Lvec \Sigmavec_{0k} \Lvec^\top) \Zvec^\top_i ).$
The EBPs are found by substituting the estimated parameters into these formulas.

Comparing the two forms of the BP, we see that the shrinkage effect i.e., the amount each predictor is shrunk toward the mean of the random effects distribution, is impacted by misspecification. 
For the true LMM, each of the terms $\mvec_k$ in \eqref{eq:BLUP_true} has two components, with one component being a scaled version of $(\Lvec \Sigmavec_{0k} \Lvec^\top)^{-1} \Lvec \muvec_{0k}$ which shrinks towards the \emph{mean of the individual mixture component} in the true random effects mixture distribution i.e.,  $\Lvec \muvec_{0k}$, and 
the other component is a scaled version of $\tau^{-2} \Zvec^\top_i (\yvec_i - \Xvec_i \betavec)$ which behaves more similarly to \eqref{eq:BLUP_mis}. It follows that this BP, being a weighted sum of the terms $\mvec_k$, $k = 1, \dots, c$, is not necessarily shrunk toward zero.
On the other hand, the BP for the misspecified LMM in \eqref{eq:BLUP_mis} is a special case of \eqref{eq:BLUP_true}, where the mixture has $c = 1$ component. This is always shrunk toward the mean of this component i.e., zero. 
This difference in shrinkage effect is less prominent if the term $\tau^{-2} \Zvec^\top_i (\yvec_i - \Xvec_i \betavec)$ is relatively large, as can occur when the cluster size is large.

\subsection{Mean squared prediction errors}\label{sec:lmm_msep}
We now turn to comparing the term $\Uvec_1$ in the UMSEP \eqref{eq:umsep} and the term $\Cvec_1$ in the CMSEP \eqref{eq:cmsep}, under the misspecified and true LMMs.

Under the misspecified LMM, the term $\Uvec_1$ is straightforwardly shown to be $\{ (\Lvec \Lvec^\top)^{-1} + \tau^{-2} \Zvec^\top_i \Zvec_i \}^{-1}$. This contrasts with the true LMM for which $\Uvec_1$ is given by
$
    \E[(\wvec_i - \uvec_i)(\wvec_i - \uvec_i)^\top] = \E[\E[(\wvec_i - \uvec_i)(\wvec_i - \uvec_i)^\top \mid \yvec_i]] = \E[\vvec_{0i}],
$
where 
$\vvec_{0i}(\thetavec) = \E[(\wvec_i - \uvec_i)(\wvec_i - \uvec_i)^\top \mid \yvec_i] = \Var[\uvec_i \mid \yvec_i] =  \{ \sum_{k=1}^c \pi_{0k} p_k(\yvec_i) \}^{-1}\sum_{k=1}^c \pi_{0k} p_k(\yvec_i) (\vvec_k + \mvec_k \mvec^\top_k) - \wvec_{0i} \wvec^\top_{0i},
$
and $\vvec_k = \{ (\Lvec \Sigmavec_{0k} \Lvec^\top)^{-1} + \tau^{-2} \Zvec^\top_i \Zvec_i \}^{-1} $. As demonstrated in the numerical study in Supplementary Material S2, the UMSEPs, which are dominated by the term $\Uvec_1$ for our simulation designs, are consistently larger under the misspecified LMM than under the true LMM, although the differences diminish as the cluster size increases.

Turning to the CMSEP,
from \eqref{eq:BLUP_mis} the difference $\wvec_{*i} - \uvec_i$ under the misspecified LMM is
\begin{align}\label{eq:lmm_pred_gap_mis}
\wvec_{*i} - \uvec_i &= [ \{ (\Lvec \Lvec^\top)^{-1} + \tau^{-2} \Zvec^\top_i \Zvec_i \} ^{-1} \tau^{-2} \Zvec^\top_i \Zvec_i - \Ivec ] \uvec_i + \{ (\Lvec \Lvec^\top)^{-1} + \tau^{-2} \Zvec^\top_i \Zvec_i \}^{-1} \tau^{-2} \Zvec^\top_i \epsilonvec_i \nonumber \\
&\triangleq \bm{\Gamma}_1 \uvec_i + \bm{\Gamma}_2 \epsilonvec_i
\end{align}
from which it follows that $\Cvec_1 = \bm{\Gamma}_1 \uvec_i \uvec^\top_i \bm{\Gamma}^\top_1 + \tau^2 \bm{\Gamma}_2 \bm{\Gamma}^\top_2$. We thus see that the CMSEP, which is dominated by $\Cvec_1$ under the setting when $m$ is large relative to $n_i$, is a quadratic function of $\uvec_i$ centered about zero for the misspecified LMM.
This can be problematic if the true random effects distribution is skewed or multimodal, as under such cases there may be potentially a non-negligible number of random effects far from zero; see Section \ref{sec:sim} for an example of this in the case of GLMMs.
By contrast, using \eqref{eq:BLUP_true} we can show the same difference under the true LMM is
\begin{equation}\label{eq:lmm_pred_gap_true}
\wvec_{0i} - \uvec_i = \sum_{k=1}^c \bm{\zeta}_k \{ ( \tau^{-2} \Zvec^\top_i \Zvec_i) \uvec_i + (\Lvec \Sigmavec_{0k} \Lvec^\top)^{-1} \Lvec \muvec_{0k} +  \tau^{-2} \Zvec^\top_i \epsilonvec_i \} - \uvec_i,
\end{equation}
where $\bm{\zeta}_k = [ \pi_{0k} p_k(\yvec_i) \{ (\Lvec \Sigmavec_{0k} \Lvec^\top )^{-1} + \tau^{-2} \Zvec^\top_i \Zvec_i \}^{-1} ] \{ \sum_{k=1}^c \pi_{0k} p_k(\yvec_i) \}^{-1}$. It follows that the form of $\Cvec_1$ and hence the CMSEP resembles a weighted quadratic function of $\uvec_i$: when the random effect is closer to the mean of mixture component $k$, the weight corresponding to that component will dominate and the overall CMSEP will be approximately equal to the quadratic function centered about $\Lvec \muvec_{0k}$. This guarantees the CMSEP to stay small around the mean of each component, in contrast to the single zero-centered quadratic form of the CMSEP under the misspecified LMM; see Supplementary Material S2 for empirical evidence of this contrasting behavior.

\section{Simulation studies} \label{sec:sim}
We performed a numerical study to compare point prediction and prediction inference under a misspecified normal random effects distribution with these under a correctly specified mixture of normal distributions, for the cases of Bernoulli and Poisson GLMMs. Note simulations for LMMs were also performed; these are available in Supplementary Material S2, and confirm some of the analytical conclusions obtained in the preceding section.

\subsection{Bernoulli response}\label{sec:sim_ber}
In the first setting, we consider a binary logistic GLMM i.e., the conditional distribution of the response follows a Bernoulli distribution with $g(\cdot)$ set to the canonical logit function, with $q_f = 2$ fixed effects and a single random intercept i.e., $q_r = 1$. We generate the elements $x_{ij2}$ from a uniform distribution $[-5,5]$, and set the fixed effects coefficients to $\betavec = (0, 1)^\top$. 
As the true distribution for the random intercept, we consider: (I) $p_0(u_i \mid \sigmavec_0) = 0.9 \N(-0.28, 0.28^2) + 0.1 \N(2.56, 1.42^2)$, and (II) $p_0(u_i \mid \sigmavec_0) = 0.5 \N(-1.77, 0.59^2) + 0.5 \N(1.77, 1.18^2)$. These two finite mixture of normal distributions were chosen to represent skewness and multimodality, noting the true parameters were chosen to satisfy the mean zero constraints and to set the variance of the random intercept equal to 1 and 4 for the two distributions, respectively.

We begin by investigating the impact of cluster size $n_i$ and number of clusters $m$ on UMSEP, by simulating datasets with $n_i = 20, 40, 60, 80$, and $m = 100, 200$.
For each combination of $m$ and $n_i$, we generate 100 data sets by sampling the random effects $\{u_i, i = 1, \dots, m\}$ from one of the two true mixture distributions, noting each dataset has a different set of random effects, and then simulating the responses $y_{ij}$ based on the set up above.
For each dataset, we then fit a binary logistic GLMM assuming either a misspecified normal distribution, or a correct two-component mixture of normal distributions, for the random effects. The fitted parameters are then substituted into equation \eqref{eq:BP} to obtain the EBPs of the random intercept for the misspecified GLMM i.e., $\hat{w}_{*i}$, and the true GLMM i.e., $\hat{w}_{0i}$. We also simulate the UMSEP by averaging the squared differences between the EBPs and the true random effect across the 100 data sets i.e., 
$
\text{UMSEP}_{\text{sim}} = 100^{-1} \sum_{s = 1}^{100} \{m^{-1} \sum_{i=1}^{m} (\hat w^{(s)}_{i} - u^{(s)}_{i})^2\}
$
where $u^{(s)}_{i}$ denotes the simulated random intercept for the $s$th dataset, and $\hat w^{(s)}_{i}$ is either $\hat{w}_{*i}$ or $\hat{w}_{0i}$ for the $s$th dataset.
The results show that, as expected, $\text{UMSEP}_{\text{sim}}$ under the misspecified GLMM is generally larger compared with the true GLMM (Table \ref{tbl:ber_glmm_umsep}). The only exception was for random effects distribution I when $n_i = 20$, and indeed we note that the difference in $\text{UMSEP}_{\text{sim}}$ between the two models is small when $m = 200$ at the smallest cluster size. We conjecture that this might be due to the fact that for a binary GLMM, estimation of a mixture of normal random effects distributions is relatively challenging and the uncertainty in the estimation contributes a non-negligible amount to the UMSEP.

\begin{table}[tb]
	\centering
	\caption{Simulated UMSEPs ($\text{UMSEP}_{\text{sim}}$) for the misspecified GLMM/true GLMM, in the case of binary responses and for two true mixture of normal random effects distributions. 
 }
	\label{tbl:ber_glmm_umsep}
	\bgroup
	\begin{tabular}{ ccccc }
		\toprule[1.5pt]
		      & $n_i = 20$ & $n_i = 40$ & $n_i = 60$ & $n_i = 80$ \\
		\cmidrule{2-5}
            & \multicolumn{4}{c}{Random effects distribution I} \\
            $m = 100$ & 0.3620/0.4289 & 0.2237/0.1507 & 0.1591/0.0996 & 0.1251/0.0776 \\
            $m = 200$ & 0.3739/0.3641 & 0.2259/0.1160 & 0.1609/0.0848 & 0.1234/0.0704 \\\\
            & \multicolumn{4}{c}{Random effects distribution II} \\
            $m = 100$ & 0.5298/0.4595 & 0.2955/0.2470 & 0.2176/0.1876 & 0.1643/0.1429 \\
\            $m = 200$ & 0.5057/0.4391 & 0.2680/0.2295 & 0.1958/0.1717 & 0.1456/0.1288 \\
                        \bottomrule[1.5pt]
	\end{tabular}
	\egroup
\end{table}

Next, we compare the CMSEP under the misspecified and true random effects distributions by fixing the cluster size to be $n_i = 40$ and the number of clusters to be $m = 400$; we choose $n_i \ll m$, so $\Cvec_1$ dominates the CMSEP. The remainder of the simulation setup is the same as above, with one crucial modification being that instead of simulating a new set of random effects for each dataset, we only simulate one set of random intercepts, and then simulate 100 data sets conditioned on this set. We then simulate the CMSEP by averaging the squared differences between the EBP and the true random effects across the 100 data sets, conditional on the single set of true random effects,
$
\text{CMSEP}_{\text{sim}} = 100^{-1} \sum_{s = 1}^{100} ( \hat{w}^{(s)}_{i} - u_{i})^2.    
$

\begin{figure}[tb]
\centering
\includegraphics[width = \textwidth]{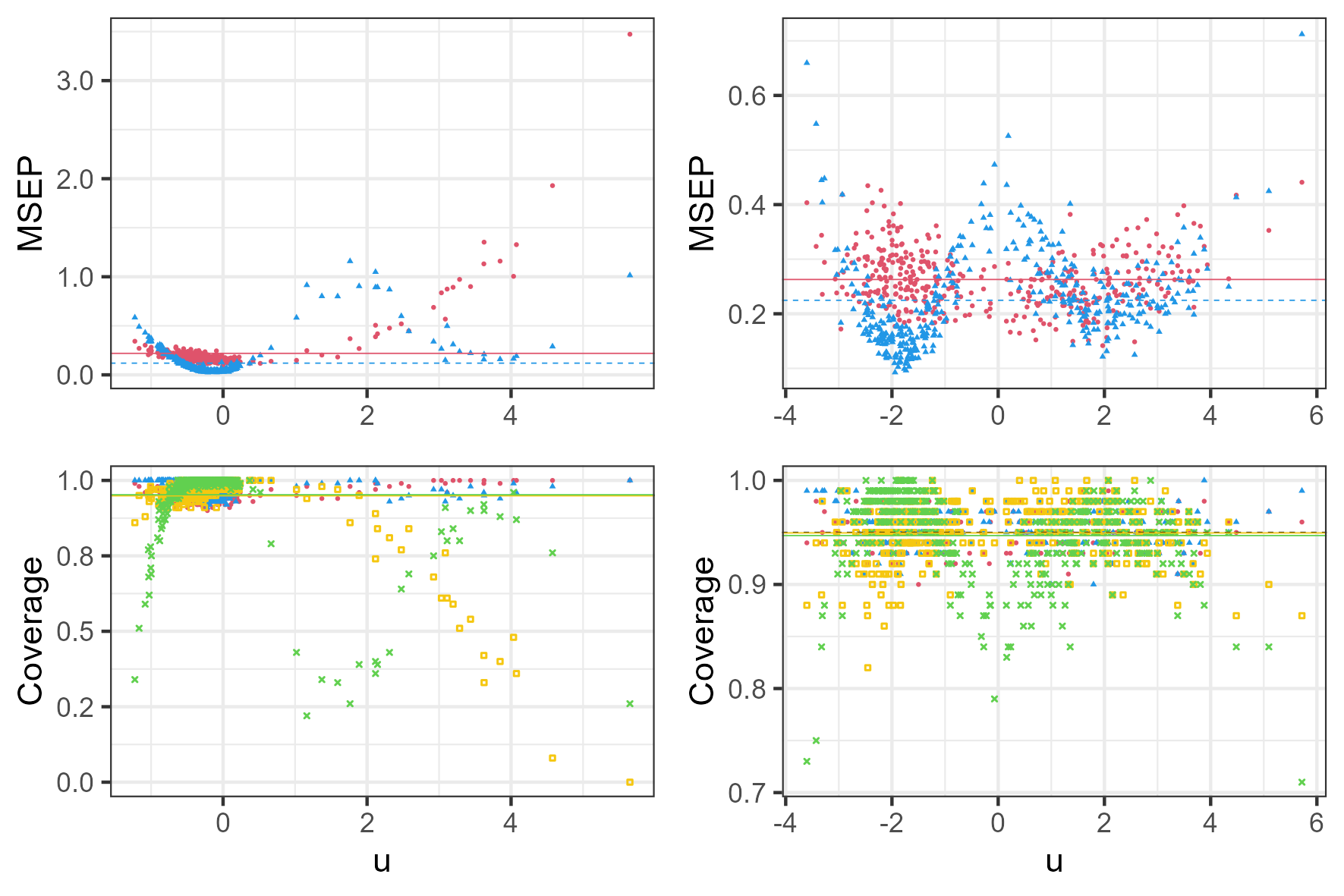}
\caption{Simulated CMSEPs ($\text{CMSEP}_{\text{sim}}$; top row) and empirical coverage probability of 95\% prediction intervals (bottom row) for the misspecified versus true GLMMs in the case of binary responses,  and for two true mixture of normal random effects distributions (left column is distribution I; right column in distribution II). 
In the top panel, the red dots and blue triangles correspond to the misspecified and true GLMMs, while in the bottom panel the red dots and blue triangles correspond to intervals constructed using $\text{CMSEP}_{\text{sim}}$ under the misspecified and true GLMMs, while the gold squares and green crosses correspond to intervals constructed using $\text{UMSEP}_{\text{sim}}$ under the misspecified and true GLMMs. 
}
\label{fig:ber_glmm_cmsep}
\end{figure}

Under the misspecified GLMM, the simulated CMSEPs are lowest around zero i.e., the assumed mean of the random effects distribution, and increase in value in the tail of the distribution (Figure \ref{fig:ber_glmm_cmsep} top row). By contrast, under the GLMM with the random effects distributions I and II, we observe two local minima for $\text{CMSEP}_{\text{sim}}$, each one at the mean of a component of the corresponding true mixture distribution. This is consistent with the analytical results for LMMs in Section \ref{sec:lmm_msep}, and shows that as we get closer to the mean of each of the mixture components the value of $\text{CMSEP}_{\text{sim}}$ produced by the misspecified GLMM is larger than under the true GLMM. In Supplementary Material S2, we display results where we decompose the simulated CMSEP into the simulated expected squared bias and the simulated variance term. Results of these show the variance terms are larger under the misspecified model especially in regions close to the mean of each of the mixture components. On the other hand, the bias under the misspecified model is close to a linear function whose absolute value gets larger when the random effects move away from zero, while the bias under the correctly specified mixture distribution appears to be a weighted linear function; the absolute bias can be a little larger under the mixture distribution at some regions of the random effects.

Finally, we examine the performance of prediction intervals for the random intercept constructed using MSEP. Following \citet{ha2011frailty, cantoni2017random,korhonen2023fast} among others, if we assume the difference $\hat w_i - u_i$ follows a normal distribution then a $100(1-\alpha)\%$ prediction interval for $u_i$ is given by
$
\hat \wvec_i \pm \Phi^{-1}\left(1-\alpha/2\right) \times (\widehat{\text{MSEP}})^{1/2},
$
where $\Phi^{-1}(\cdot)$ denotes the inverse cumulative distribution function of the normal distribution and $\widehat{\text{MSEP}}$ is either the simulated UMSEP or simulated CMSEP. Note in practice we do not observe the true random effect, and so we need to estimate the MSEP e.g., using the parametric bootstrap approach discussed at the end of Section \ref{sec:msep}.

The results show 95\% prediction intervals based on the simulated UMSEP under both the true and misspecified GLMMs approximately achieve the nominal marginal coverage probability i.e., when averaged over all the random effects values (Figure \ref{fig:ber_glmm_cmsep} bottom panel). This is because the UMSEP under the misspecified distribution still correctly captures the uncertainty of the BP. Note we do not present results for interval widths here, since these are equal to a scaled version of the simulated UMSEP/CMSEP, where the scaling factor is constant regardless of $u_i$. 
Turning to the coverage probability conditional on specific values of $u_i$, 
generally 95\% prediction intervals constructed using simulated CMSEPs achieve reasonable coverage regardless of the true value of $u_i$ and for both true mixture of normal random effects distributions. By contrast, the conditional coverage probability for prediction intervals constructed using the simulated UMSEPs is less than the nominal coverage for certain regions of the random effects distribution e.g., at the right tail of distribution I. This is a direct consequence of the UMSEP being lower than the CMSEP at a certain value of the random effect, and is not too surprising given UMSEP (and hence prediction intervals constructed using it) is based on the idea of averaging across all the random effects i.e., there is no guarantee it will be correct conditional on a specific value of $u_i$.

\subsection{Poisson response} \label{subsec:poissonglmm}
In the second setting, we consider a Poisson GLMM i.e., the conditional distribution of the response follows a Poisson distribution with the canonical log link. We again used $q_f = 2$ fixed effects with the first set to an intercept term and $x_{ij2}$ drawn independently from the standard uniform distribution, $\betavec = (0, 1)^\top$, and a single random intercept. As the true random effects distributions, we investigate two mixtures of normal distributions similar to the binary GLMM case. The remainder of the simulation setup and performance assessment was similar to the binary response case in Section \ref{sec:sim_ber}. 

Given the availability of more information in the response compared to the binary response in Section \ref{sec:sim_ber}, we consider smaller values of $n_i = 5, 10, 20, 40$ and $m = 50, 100$ and investigate their impact on UMSEP. The results show that, similar to the binary response case, larger values of $\text{UMSEP}_{\text{sim}}$ occur under the misspecified compared with the true GLMM (Table \ref{tbl:pois_glmm_umsep}), with the differences being more evident when the cluster size is small.

\begin{table}[tb]
	\centering
	\caption{Simulated UMSEPs ($\text{UMSEP}_{\text{sim}}$) for the misspecified GLMM/true GLMM, in the case of Poisson responses and for two true mixture of normal random effects distributions. 
 }
	\label{tbl:pois_glmm_umsep}
	\bgroup
	\begin{tabular}{ ccccc }
		\toprule[1.5pt]
		      & $n_i = 5$ & $n_i = 10$ & $n_i = 20$ & $n_i = 40$ \\
		\cmidrule{2-5}
            & \multicolumn{4}{c}{Random effects distribution I} \\
            $m = 50$ & 0.1458/0.0836 & 0.0838/0.0607 & 0.0551/0.0458 & 0.0400/0.0371 \\
            $m = 100$ & 0.1434/0.0682 & 0.0782/0.0496 & 0.0436/0.0330 & 0.0264/0.0235 \\\\
            & \multicolumn{4}{c}{Random effects distribution II} \\
            $m = 50$ & 0.4062/0.2665 & 0.2790/0.1900 & 0.1725/0.1244 & 0.1426/0.1271 \\
\            $m = 100$ & 0.3543/0.2100 & 0.2284/0.1364 & 0.1569/0.1122 & 0.0997/0.0841 \\
                        \bottomrule[1.5pt]
	\end{tabular}
	\egroup
\end{table}

\begin{figure}[htb]
\centering
\includegraphics[width = \textwidth]{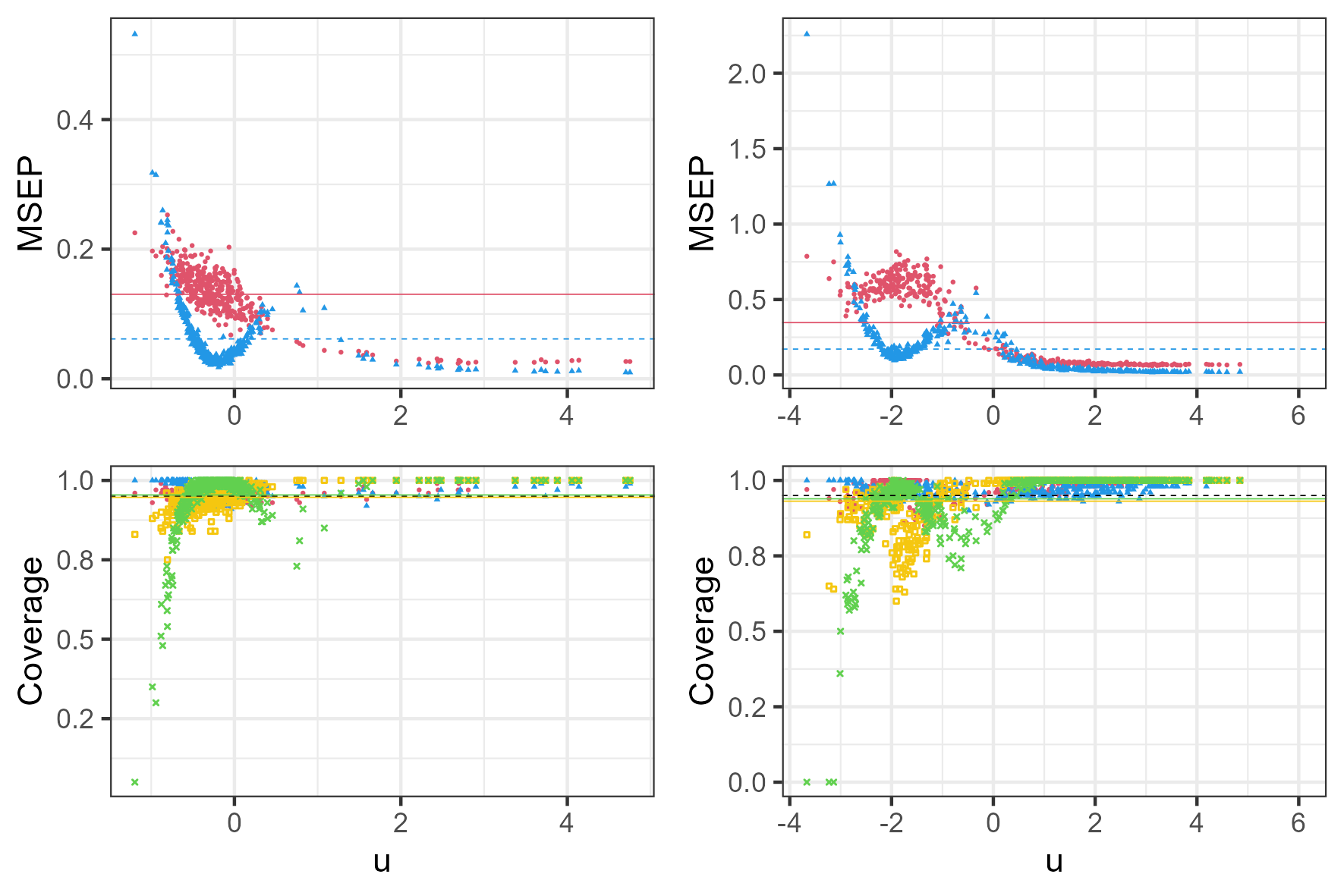}
\caption{Simulated CMSEPs ($\text{CMSEP}_{\text{sim}}$; top row) and empirical coverage probability of 95\% prediction intervals (bottom row)  for misspecified versus true GLMMs in the case Poisson responses, and for two true mixture of normal random effects distributions (left column is distribution I; right column is distribution II). 
In the top panel, the red dots and and blue triangles correspond to the misspecified and true GLMMs, while in the bottom panel the red dots and blue triangles correspond to intervals constructed using $\text{CMSEP}_{\text{sim}}$ under the misspecified and true GLMMs, while the gold squares and green crosses correspond to intervals constructed using $\text{UMSEP}_{\text{sim}}$ under the misspecified and true GLMMs. 
}
\label{fig:pois_glmm_cmsep}
\end{figure}

Turning to comparisons of CMSEP, we fix the cluster size at $n_i = 5$ and the number of clusters at $m = 400$. For both true random effects distributions I and II, the true GLMM produces considerably smaller values of the simulated CMSEP around the mean of the first component in the mixture distribution (Figure \ref{fig:pois_glmm_cmsep}). Around the mean of the second mixture component, the true GLMM also returns smaller simulated CMSEPs compared to its misspecified counterpart, although the differences are smaller. This shows that the link function in the GLMM can impact the behavior of $\text{CMSEP}_{\text{sim}}$ i.e., using the log link the CMSEPs for larger random effect values are small and tending to zero.
In Supplementary Material S2 we show results decomposing the simulated CMSEP into a bias and variance term. Similar to Section \ref{sec:sim_ber}, these results show the variances are larger under the misspecified model, especially around the mean of the first mixture component, while the absolute bias can be larger for the correctly specified GLMM in some regions of the random effects distribution.

Finally, the results for the marginal and conditional coverage probability are largely similar to those for the binary GLMM case (Figure \ref{fig:pois_glmm_cmsep}). In particular, the 95\% prediction intervals constructed using simulated CMSEPs achieve reasonable conditional coverage probabilities. For marginal coverage probabilities, the intervals constructed with simulated UMSEPs perform relatively well.

\section{Case study: Longitudinal survey of hourly wages }\label{sec:case_study}
To illustrate the above analytical and empirical findings in a real application, we analyze data from the National Longitudinal Survey of Youth \citep{singer2003applied} containing the log of hourly wages for $m = 888$ individuals, each measured at $n_i$ occasions, with cluster size  $n_i$ ranging from 1 to 13, $i=1,\ldots,m$. We fitted an independent cluster LMM where time spent in the workforce, race, and education along with an intercept were included as fixed effect covariates (so $q_f = 4$), and a single $q_r = 1$ random intercept is included to account for between-individual heterogeneity. 

\begin{figure}[tb]
\centering
\includegraphics{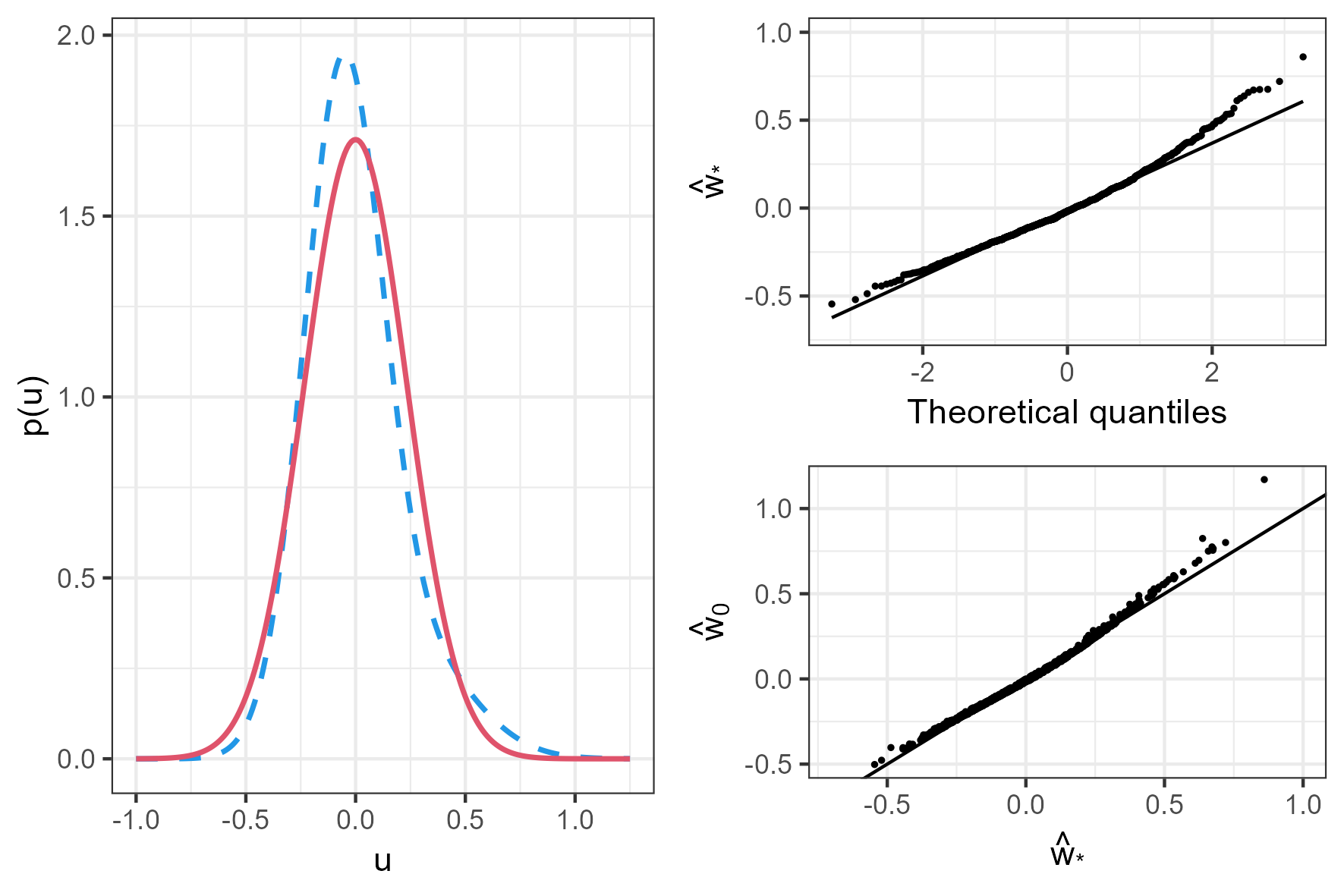}
\caption{Comparison of the random effects under the normal and mixture of normals LMMs fitted for the wages data. Left panel: the fitted random effects distribution (red solid line for normal LMM and blue dashed line for mixture of normals LMM). Right panel: the EBPs under the normal distribution against theoretical quantiles of the normal distribution (top), and against the EBPs under the mixture distribution (bottom).}
\label{fig:wages1}
\end{figure}

We begin by fitting the standard LMM assuming the random effects are normally distributed. 
A check of the resulting predicted random effects using both a normal quantile-quantile plot and a Shapiro-Wilk test at the 5\% significance level offers statistical evidence of a deviation from normality, indicating that the assumed normality does not hold. In particular, there is evidence that the distribution may be right-skewed (see top right panel of Figure \ref{fig:wages1}).
With this in mind, we proceed to fit an LMM using a two-component mixture of normal distributions for the random intercept, as formulated above Theorem \ref{thm:LMMequivalence}.
Figure \ref{fig:wages1} presents the estimated random effects distributions along with the EBPs for the two LMMs fitted, from which we see that the fitted mixture of normal distributions is slightly right skewed (left panel). The EBPs from this fit are comparably less shrunk toward zero than those produced by the normal random effects LMM (bottom right panel); this result is consistent with the theoretical findings in Section \ref{sec:lmm_predictor}.

\begin{figure}[tb]
\centering
\includegraphics[width = 0.8\textwidth]{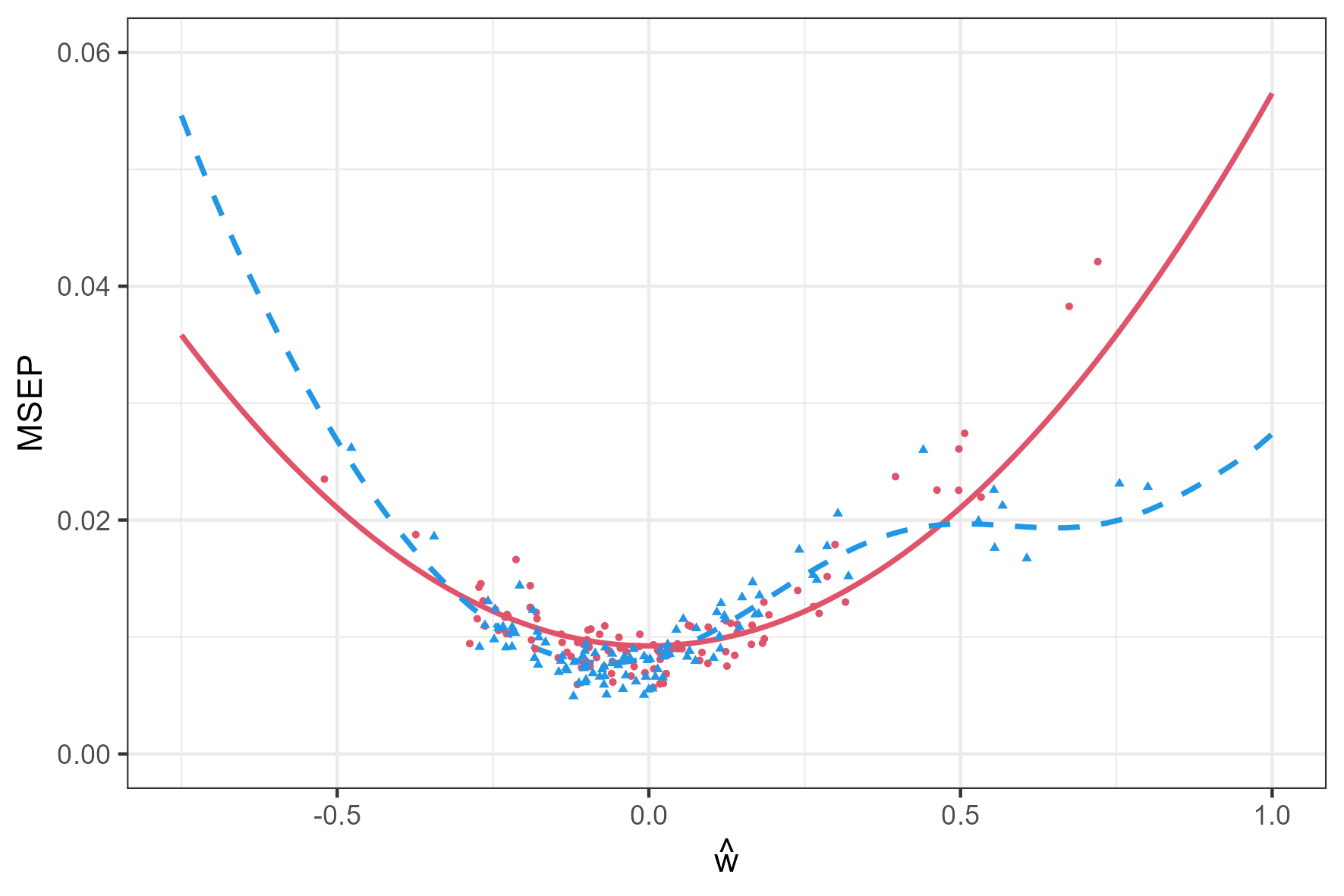}
\caption{Comparison of CMSEP under LMMs fitted for the wages data, assuming either the normal or mixture of normals random effects distribution. The curves correspond to the analytical $\Cvec_1$ term i.e., equations \eqref{eq:lmm_pred_gap_mis} (red solid line) and \eqref{eq:lmm_pred_gap_true} (blue dashed line), while 
the red dots and blue triangles are parametric bootstrap estimates of the CMSEP under the normal and mixture distributions.}
\label{fig:wages2}
\end{figure}

Next, we examine the MSEP for the two fitted LMMs. Given that the number of clusters $m$ is in the hundreds and relatively large, while the cluster sizes $n_i$ are no larger than 13 and relatively small, then as discussed in Section \ref{sec:msep} the $\Cvec_1$  term in the CMSEP is expected to dominate in equation \eqref{eq:cmsep}. 
As a result, based on equations \eqref{eq:lmm_pred_gap_mis} and \eqref{eq:lmm_pred_gap_true}, which compute the difference between the best predictor and the true random effects value, we can approximate the CMSEP under the normal and mixture of normals LMMs respectively, where the estimated parameters are substituted into the equations.
For illustrative purposes, we subset to only consider 103 individuals with cluster size $n_i = 7$ (the mean cluster size).
Similar to the results of the simulation study for LMMs (Supplementary Material S2) and for GLMMs broadly (Section \ref{sec:sim}), for many random effect values the CMSEPs under the mixture of normals random effects distribution is lower than the CMSEPs under the normal random effects distribution (Figure \ref{fig:wages2} solid and dotted curves). This is especially the case around the mean of the main component of the mixture and in the heavy right tail.
Alternatively, we can use the bootstrap to estimate the MSEPs, as discussed towards the end of Section \ref{sec:msep}. The bootstrap estimates of the CMSEPs are shown in Figure \ref{fig:wages2}. Note that in comparison to using the analytical expressions for $\Cvec_1$, the bootstrap estimates account for estimating the full MSEP e.g., incorporating terms $\Cvec_2$ and $\Cvec_3$ as well. On the other hand, because bootstrapping is performed given the EBPs of the random effects, then the bootstrap estimates must be interpreted carefully since they are naturally biased e.g., predictors under the normal assumption at the tail are shrunk toward zero more than it should be \citep[see][among others for solutions to this problem in practice]{carpenter2003novel}.
Overall, results from this parametric bootstrap approach are largely consistent with the analytical forms of $\Cvec_1$.
Moreover, the bootstrap estimate of the UMSEP under the normal distribution is 0.0112, which is slightly larger than the bootstrap estimate of the UMSEP, 0.0108 under the mixture distribution; also, the term $\Uvec_1$ is equal to 0.0134 under the normal distribution, which is larger than the term $\Uvec_1$, 0.0130 under the mixture distribution. This result is consistent with the findings in the simulation studies in Section \ref{sec:sim}.

Finally, Supplementary Material S2 also presents results of 95\% prediction intervals constructed using bootstrap estimates of the CMSEP, from which we see that at the right tail of the random effects distribution, the intervals under the normal LMM are much wider. This may suggest that misspecifying the random effects distribution (assuming the mixture of normals LMM is a more accurate representation of the true data generation process) could lead to conclusions with higher prediction uncertainty.

\section{Discussion}\label{sec:conclusion}
In this article, we studied the effect of misspecifying the random effects distribution on prediction of random effects in GLMMs. We compared point prediction and prediction uncertainty under a misspecified normal distribution versus a correctly specified mixture of normal distributions.
The mathematical derivations for the case of LMMs along with empirical findings for the case of GLMMs result in a number of key findings.  
First, the degree of shrinkage for random effects is affected: while the BPs and EBPs under a misspecified normal distribution are shrunk toward zero, the corresponding predictors under the true mixture distribution are shrunk differently and not necessarily toward zero. Practically, this may lead to excessively large bias for random effects predictions, particularly in the tail of the random effects distribution, under misspecification.
Second, MSEPs are also negatively impacted under misspecification, with UMSEPs consistently larger and the effects being most evident in settings with small cluster sizes. The case of CMSEPs is more complicated, although they still tend to remain larger under the misspecified normal distribution, especially those in the regions of the mean of a component of the underlying mixture distribution.
Finally, prediction intervals based on MSEPs are wider under the misspecified normal distribution. While prediction intervals using CMSEPs achieve nominal coverage across clusters, conditional coverage probabilities of intervals based on UMSEPs vary across clusters.

While this article has focused on just the consequences of misspecification on prediction of random effects alone, predicting a function $t(\uvec_i)$ of the random effect $\uvec_i$ is also often of interest e.g., in fisheries ecology one might need to predict the expected number of (non-zero) bycatch, which is a complicated non-linear function of the fixed and random effects \citep{cantoni2017random}.
We can use the (empirical) best predictor for prediction of this function \citep[e.g.,][]{vidoni2006response, skrondal2009prediction}, such that the point predictor for $t(\uvec_i)$ is given by is 
$
\E[t(\uvec_i) \mid \yvec_i] = \int t(\uvec_i) p(\uvec_i \mid \yvec_i) \txd\uvec_i.
$
We conjecture that the overall conclusions of this article will likely carry over to prediction inference of functions of random effects, and once again misspecification is likely to have negative impacts on aspects such as shrinkage and conditional coverage probabilities. 
Also, this article has focused on the (empirical) best predictors. 
It would be interesting to investigate how other predictors, such as one based on quantile regression \citep[e.g.,][]{chambers2006m}, or one based on the mode of the conditional distribution of the random effects \citep[commonly used when the Laplace approximation or penalized quasi-likelihood is applied to mixed models]{brooks2017,hui2021use}, perform under misspecification.
Finally, we focused on misspecification of the random effects distribution, when assuming other elements of the generalized linear mixed models (such as the conditional distribution of the responses, the link function, etc) are correctly specified. It would be of interest to study the impact of (joint) misspecification of these elements on point prediction and prediction inference.

\section*{Acknowledgements}
This work was supported by the Australian Research Council under Grant DP230101908. Thank you to Nickson Xu Ning for useful discussions.

\bibliographystyle{apalike}
\bibliography{biblio}

\newpage    
\appendix
\def\thesection{S\arabic{section}}
\def\thefigure{S\arabic{figure}}
\def\thetable{S\arabic{table}}

\section{Proofs and Derivations}\label{sec:appendix}

\subsection{Proof of Theorem 1}
Suppose the response $\yvec = (\yvec^\top_1, \dots, \yvec^\top_m)^\top$ is modeled as
$$
\yvec = \Xvec \betavec + \Zvec \uvec + \epsilonvec,
$$
where 
$\Xvec = (\Xvec^\top_1, \dots, \Xvec^\top_m)^\top, \Zvec = \bdiag(\Zvec^\top_1, \dots, \Zvec^\top_m), \uvec = (\uvec^\top_1, \dots, \uvec^\top_m)^\top$. As $p_0(\uvec_i) = \sum_{k=1}^c \pi_{0k} \N(\Lvec_0 \muvec_{0k}, \Lvec_0 \Sigmavec_{0k} \Lvec_0^\top)$, $p_{*}(\uvec_i) = \N(0, \Sigmavec_{*})$, and $\Lvec_{*} = \text{chol}(\Sigmavec_{*})$ for $i = 1, \dots, m$, then
the true joint distribution of $\uvec$ is a mixture of normal distributions, that is, $p_0(\uvec) = \sum_{\tilde k=1}^{\tilde c} \tilde{\pi}_{\tilde k} \N(\tilde{\Lvec}_0 \tilde{\muvec}_{0 \tilde k}, \tilde{\Lvec}_0 \tilde{\Sigmavec}_{0 \tilde k} \tilde{\Lvec}^\top_0)$, with the constraints $\sum_{\tilde k=1}^{\tilde c} \tilde{\pi}_{\tilde k} \tilde{\muvec}_{0 \tilde k} = \zerovec$, and $\sum_{\tilde k=1}^{\tilde c} \tilde{\pi}_{\tilde k} (\tilde{\Sigmavec}_{0 \tilde{k}} + \tilde{\muvec}_{0 \tilde k} \tilde{\muvec}^\top_{0 \tilde k}) = \Ivec$, and $\tilde{\Lvec}_0 = \bdiag((\Lvec_0)_m)$. 
The misspecified joint distribution of $\uvec$ is $p_{*}(\uvec) = \N(\zerovec, \tilde{\Lvec}_{*} \tilde{\Lvec}^\top_{*})$, where $\tilde{\Lvec}_{*} = \bdiag((\Lvec_{*})_m)$. 

Under the true model, $p(\yvec \mid \uvec) = \N(\Xvec \betavec_0 + \Zvec \uvec, \tau_0^2 \Ivec)$, while under the misspecified model, $p(\yvec \mid \uvec) = \N(\Xvec \betavec_{*} + \Zvec \uvec, \tau_{*}^{2} \Ivec)$.
Then, the marginal distribution of $\yvec$ under the true model is $p_0(\yvec) = \sum_{\tilde k=1}^{\tilde c} \tilde{\pi}_{\tilde k} \N(\Xvec \betavec_0 + \Zvec \tilde{\Lvec}_0 \tilde{\muvec}_{0 \tilde k}, \Zvec \tilde{\Lvec}_0 \tilde{\Sigmavec}_{0 \tilde k} \tilde{\Lvec}^\top_0 \Zvec^\top + \tau_0^2 \Ivec) = \sum_{\tilde k=1}^{\tilde c} \N(\tilde{\muvec}_{\tilde k}, \tilde{\Sigmavec}_{\tilde k}) = \sum_{\tilde k=1}^{\tilde c} \tilde{\pi}_{\tilde k} p_{\tilde k}(\yvec)$. 
The marginal distribution of $\yvec$ under the misspecified model is $p_{*}(\yvec) = \N(\Xvec \betavec_{*}, \Zvec \tilde{\Lvec}_{*} \tilde{\Lvec}^\top_{*} \Zvec^\top + \tau_{*}^2 \Ivec) = \N(\tilde{\muvec}_{*}, \tilde{\Sigmavec}_{*})$.

The negative Kullback-Leibler divergence between the true marginal distribution of $\yvec$ and the misspecified marginal distribution of $\yvec$, denoted $-\operatorname{KL}(\cdot || \cdot)$ is
\begin{align*}
K = - \operatorname{KL}(p_0(\yvec) || p_{*}(\yvec)) &= \int p_0(\yvec) \log(\frac{p_0(\yvec)}{p_{*}(\yvec)}) \txd\yvec \\
&= \const + \int p_0(\yvec) \log({p_{*}(\yvec)}) \txd\yvec \\
&= \const + \int \sum_{\tilde k = 1}^{\tilde c} \tilde{\pi}_{\tilde k} p_{\tilde k}(\yvec)  \log({p_{*}(\yvec)}) \txd\yvec \\
&= \const + \frac{1}{2} \sum_{\tilde k = 1}^{\tilde c} \tilde{\pi}_{\tilde k} \{\trc(\tilde{\Sigmavec}_{*}^{-1} \tilde{\Sigmavec}_{\tilde k}) + (\tilde{\muvec}_{*} - \tilde{\muvec}_{\tilde k})^\top \tilde{\Sigmavec}_{*}^{-1} (\tilde{\muvec}_{*} - \tilde{\muvec}_k) + \log\abs{\tilde{\Sigmavec}_{*}} \}.
\end{align*}

Now, we need to optimize this function to find the values of $\thetavec_{*}$.

Differentiating $K$ with respect to $\tilde{\muvec}_{*}$, we obtain
\begin{align*}
\frac{\partial K}{\partial \tilde{\muvec}_{*}} &= \sum_{\tilde k=1}^{\tilde c} \tilde{\pi}_{\tilde k} (\tilde{\muvec}_{*} - \tilde{\muvec}_{\tilde k})^\top \tilde{\Sigmavec}_{*}^{-1} \\
&= \sum_{\tilde k=1}^{\tilde c} \tilde{\pi}_{\tilde k} (\Xvec \betavec_{*} - ( \Xvec \betavec_0 + \Zvec \tilde{\Lvec}_0  \tilde{\muvec}_{0 \tilde k} ) )^\top  \tilde{\Sigmavec}_{*}^{-1} \\
&= (\Xvec \betavec_{*} - \Xvec \betavec_0)^\top \tilde{\Sigmavec}_{*}^{-1} \\
&= (\betavec_{*} - \betavec_0)^\top \Xvec^\top \tilde{\Sigmavec}_{*}^{-1}.
\end{align*}
It follows that setting $\partial K/ \partial \tilde{\muvec}_{*} = \bm{0}$ results in $\betavec_{*} - \betavec_0 = 0$.

Differentiating $K$ with respect to $\tilde{\Sigmavec}_{*}$, we obtain
\begin{align*}
\frac{\partial K}{\partial \tilde{\Sigmavec}_{*}} &= \sum_{\tilde k=1}^{\tilde c} \frac{\tilde{\pi}_{\tilde k}}{2} \left(\tilde{\Sigmavec}_k + (\tilde{\muvec}_{*} - \tilde{\muvec}_k)(\tilde{\muvec}_{*} - \tilde{\muvec}_k)^\top - \tilde{\Sigmavec}_{*} \right) \\
&= \sum_{\tilde k=1}^{\tilde c} \frac{\tilde{\pi}_{\tilde k}}{2} \left( \Zvec \tilde{\Lvec}_0  \tilde{\Sigmavec}_{0 \tilde k} \tilde{\Lvec}^\top_0 \Zvec^\top + \tau_0^2 \Ivec +  \Zvec \tilde{\Lvec}_0 \tilde{\muvec}_{0 \tilde k} \tilde{\muvec}^\top_{0 \tilde k} \tilde{\Lvec}^\top_0  \Zvec^\top - ( \Zvec \tilde{\Lvec}_{*} \tilde{\Lvec}^\top_{*} \Zvec^\top + \tau_{*}^2 \Ivec) \right) \\
&= \frac{1}{2} \Zvec \tilde{\Lvec}_0  \left[\sum_{\tilde k=1}^{\tilde c} \tilde{\pi}_{\tilde k} (\tilde{\Sigmavec}_{0 \tilde k} + \tilde{\muvec}_{0 \tilde k} \tilde{\muvec}^\top_{0 \tilde k}) \right] \tilde{\Lvec}^\top_0 \Zvec^\top - \frac{1}{2} \Zvec \tilde{\Lvec}_{*} \tilde{\Lvec}^\top_{*}  \Zvec^\top + \frac{\tau_0^2 - \tau_{*}^2}{2} \Ivec \\
&= \frac{1}{2} \Zvec (\tilde{\Lvec}_0 \tilde{\Lvec}^\top_0 - \tilde{\Lvec}_{*} \tilde{\Lvec}^\top_{*} ) \Zvec^\top + \frac{\tau_0^2 - \tau_{*}^2}{2} \Ivec.
\end{align*}
It follows that setting $\partial K/\partial \tilde{\Sigmavec}_{*} = \bm{0}$ results in $\tilde{\Lvec}_0  - \tilde{\Lvec}_{*}  = 0$ and $\tau_0^2 - \tau_{*}^2 = 0$.

\subsection{Derivation of best predictor formula for LMMs}

First, note that if $p(\yvec \mid \uvec) = \N(\Zvec \uvec + \bvec, \Tvec)$, and $p(\uvec) = \N(\muvec, \Svec)$, then $p(\yvec) = \N(\Mvec_y, \Vvec_y)$, and $p(\uvec \mid \yvec) = \N(\Mvec_u, \Vvec_u)$, where
$\Mvec_y = \bvec + \Zvec \muvec$,
$\Vvec_y = \Tvec + \Zvec \Svec \Zvec^\top$,
$\Mvec_u = (\Svec^{-1} + \Zvec^\top \Tvec^{-1} \Zvec)^{-1} (\Svec^{-1} \muvec + \Zvec^\top \Tvec^{-1} (\yvec - \bvec))$, and
$\Vvec_u = (\Svec^{-1} + \Zvec^\top \Tvec^{-1} \Zvec)^{-1}$.

Then, we can see that the best predictor $\wvec_{0i}$ of the random intercept $\uvec_i$ for cluster $i$ under the true model is given by
\begin{align*}
\wvec_{0i}(\thetavec_0) &= \E_0(\uvec_i \mid \yvec_i) = \int \uvec_i p_0(\uvec_i \mid \yvec_i) \txd \uvec_i \\
&= \frac{\int \uvec_i \sum_{k=1}^c \pi_{0k} \prod_{j=1}^{n_i} p(y_{ij}\mid\betavec, \phi, \uvec_i) p_k(\uvec_i) \txd \uvec_i}{\int \sum_{k=1}^c \pi_{0k} \prod_{j=1}^{n_i} p(y_{ij}\mid\betavec, \phi, \uvec_i) p_k(\uvec_i) \txd \uvec_i} \\
&= \frac{\int \uvec_i \sum_{k=1}^c \pi_{0k} p_k(\yvec_i) p_k(\uvec_i \mid \yvec_i) \txd \uvec_i}{\int \sum_{k=1}^c \pi_{0k} p_k(\yvec_i) p_k(\uvec_i \mid \yvec_i) \txd \uvec_i} \\
&= \frac{\sum_{k=1}^c \pi_{0k} p_k(\yvec_i) \int \uvec_i p_k(\uvec_i \mid \yvec_i) \txd \uvec_i}{\sum_{k=1}^c \pi_{0k} p_k(\yvec_i) \int p_k(\uvec_i \mid \yvec_i) \txd \uvec_i} \\
&= \frac{\sum_{k=1}^c \pi_{0k} p_k(\yvec_i) \mvec_k}{\sum_{k=1}^c \pi_{0k} p_k(\yvec_i)},
\end{align*}
where $\mvec_k = ((\Lvec \Sigmavec_{0k} \Lvec^\top)^{-1} + \tau^{-2} \Zvec^\top_i \Zvec_i)^{-1} ((\Lvec \Sigmavec_{0k} \Lvec^\top)^{-1} \Lvec \muvec_{0k} + \tau^{-2} \Zvec^\top_i (\yvec_i - \Xvec_i \betavec))$.

\subsection{Derivation of MSEP formula for LMMs}

The prediction variance under the true model is given by
\begin{align*}
\vvec_{0i}(\thetavec_0) &= \Var_0(\uvec_i \mid \yvec_i) = \E_0(\uvec_i \uvec^\top_i \mid \yvec_i) -  [\E_0(\uvec_i \mid \yvec_i)] [\E_0(\uvec_i \mid \yvec_i)]^\top \\
&= \int \uvec_i \uvec^\top_i p_0(\uvec_i \mid \yvec_i) \txd \uvec_i - \wvec_{0i} \wvec^\top_{0i}  \\
&= \frac{\int \uvec_i \uvec^\top_i \sum_{k=1}^c \pi_{0k} \prod_{j=1}^{n_i} p(y_{ij}\mid\betavec, \phi, \uvec_i) p_k(\uvec_i) \txd \uvec_i}{\int \sum_{k=1}^c \pi_{0k} \prod_{j=1}^{n_i} p(y_{ij}\mid\betavec, \phi, \uvec_i) p_k(\uvec_i) \txd \uvec_i} - \wvec_{0i} \wvec^\top_{0i}  \\
&= \frac{\int \uvec_i \uvec^\top_i \sum_{k=1}^c \pi_{0k} p_k(\yvec_i) p_k(\uvec_i \mid \yvec_i) \txd \uvec_i}{\int \sum_{k=1}^c \pi_{0k} p_k(\yvec_i) p_k(\uvec_i \mid \yvec_i) \txd \uvec_i} - \wvec_{0i} \wvec^\top_{0i} \\
&= \frac{\sum_{k=1}^c \pi_{0k} p_k(\yvec_i) \int \uvec_i \uvec^\top_i p_k(\uvec_i \mid \yvec_i) \txd \uvec_i}{\sum_{k=1}^c \pi_{0k} p_k(\yvec_i) \int p_k(\uvec_i \mid \yvec_i) \txd \uvec_i} - \wvec_{0i} \wvec^\top_{0i} \\
&= \frac{\sum_{k=1}^c \pi_{0k} p_k(\yvec_i) (\vvec_k + \mvec_k \mvec^\top_k)}{\sum_{k=1}^c \pi_{0k} p_k(\yvec_i)} - \wvec_{0i} \wvec^\top_{0i},
\end{align*}
where $\vvec_k = ((\Lvec \Sigmavec_{0k} \Lvec^\top)^{-1} + \tau^{-2} \Zvec^\top_i \Zvec_i)^{-1} $.

\subsection{Parametric bootstrap algorithm}

\begin{algorithm}
Obtain the maximum likelihood estimate $\hat \thetavec$, and empirical best predictor $\hat \wvec_i$. For $b = 1, \dots, B$ do \\
\nl 
To approximate the UMSEP, generate bootstrap samples for the random effects: $\uvec_{i, b} \sim p(\uvec; \hat \thetavec)$. \\
To approximate the CMSEP, fix the random effects to the predicted random effects: $\uvec_{i, b} = \hat \wvec_i$. \\
\nl Generate bootstrap samples $y_{ij, b} \sim p(y_{ij} \mid \uvec_{i, b}; \hat \thetavec)$. \\
\nl Obtain the bootstrap estimates $\hat \thetavec_b$, and empirical best predictor $\hat \wvec_{i, b}$. \\
The MSEPs can then be approximated from the bootstrap predictors, that is,
\begin{align*}
\hat{\text{MSEP}} = \E[(\hat \wvec_i - \uvec_i)(\hat \wvec_i - \uvec_i)^\top] \approx \frac{1}{B} \sum_{b=1}^B (\hat \wvec_{i, b} - \uvec_{i, b})(\hat \wvec_{i, b} - \uvec_{i, b})^\top.
\end{align*}
\caption{{\bf Parametric bootstrap} \label{alg:parametric_boot}}
\end{algorithm}

\newpage
\section{Additional results} \label{app:sims_moreresults}

\subsection{Additional figures}

\begin{figure}[h]
\centering
\includegraphics{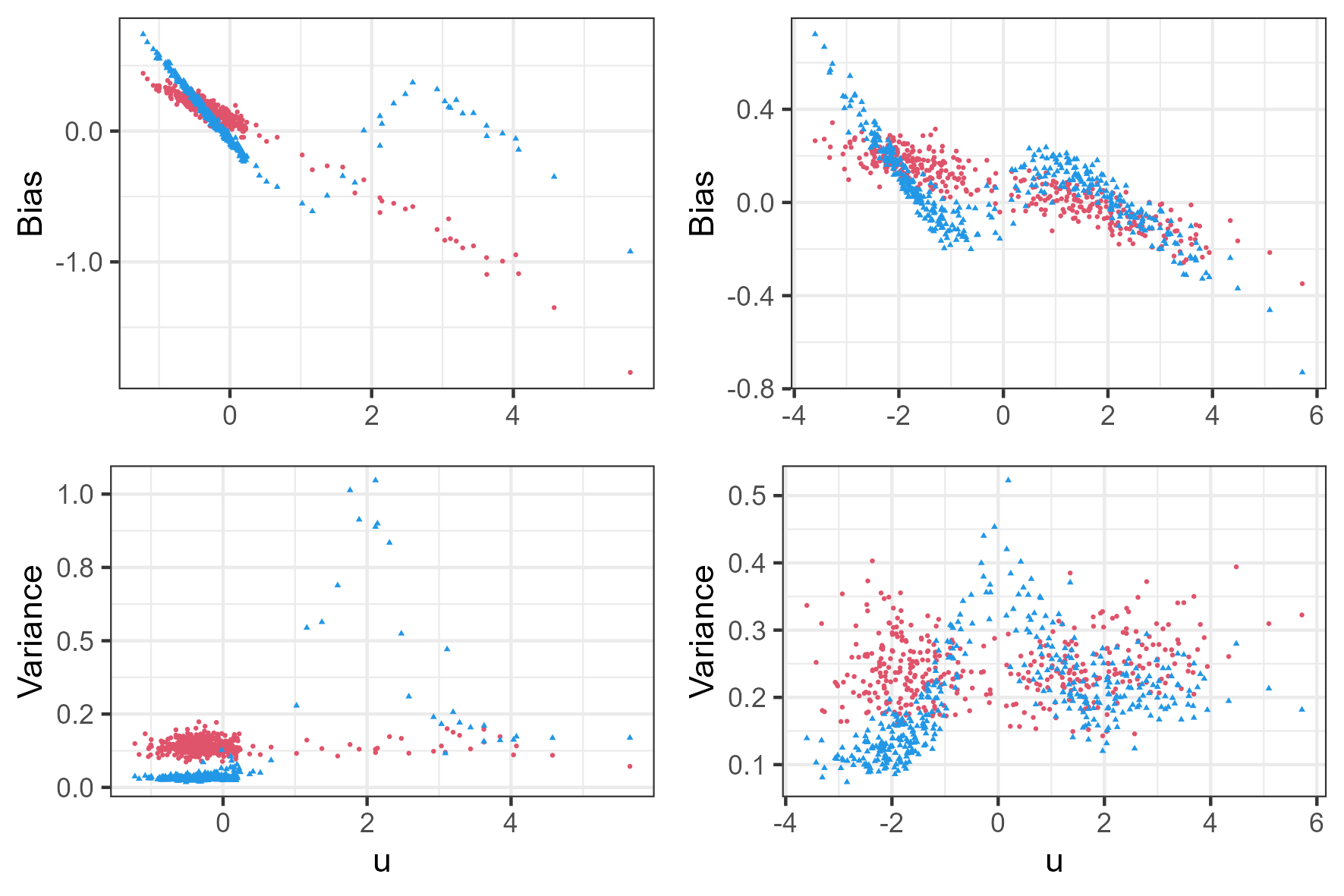}
\caption{
Bias and variance decomposition of the simulated CMSEP for the Bernoulli GLMM simulations. Red: Under the misspecified distribution. Blue: Under the correctly specified distribution. Left panel: True distribution is Distribution 1. Right panel: True distribution is Distribution 2.}
\end{figure}

\begin{figure}[h]
\centering
\includegraphics{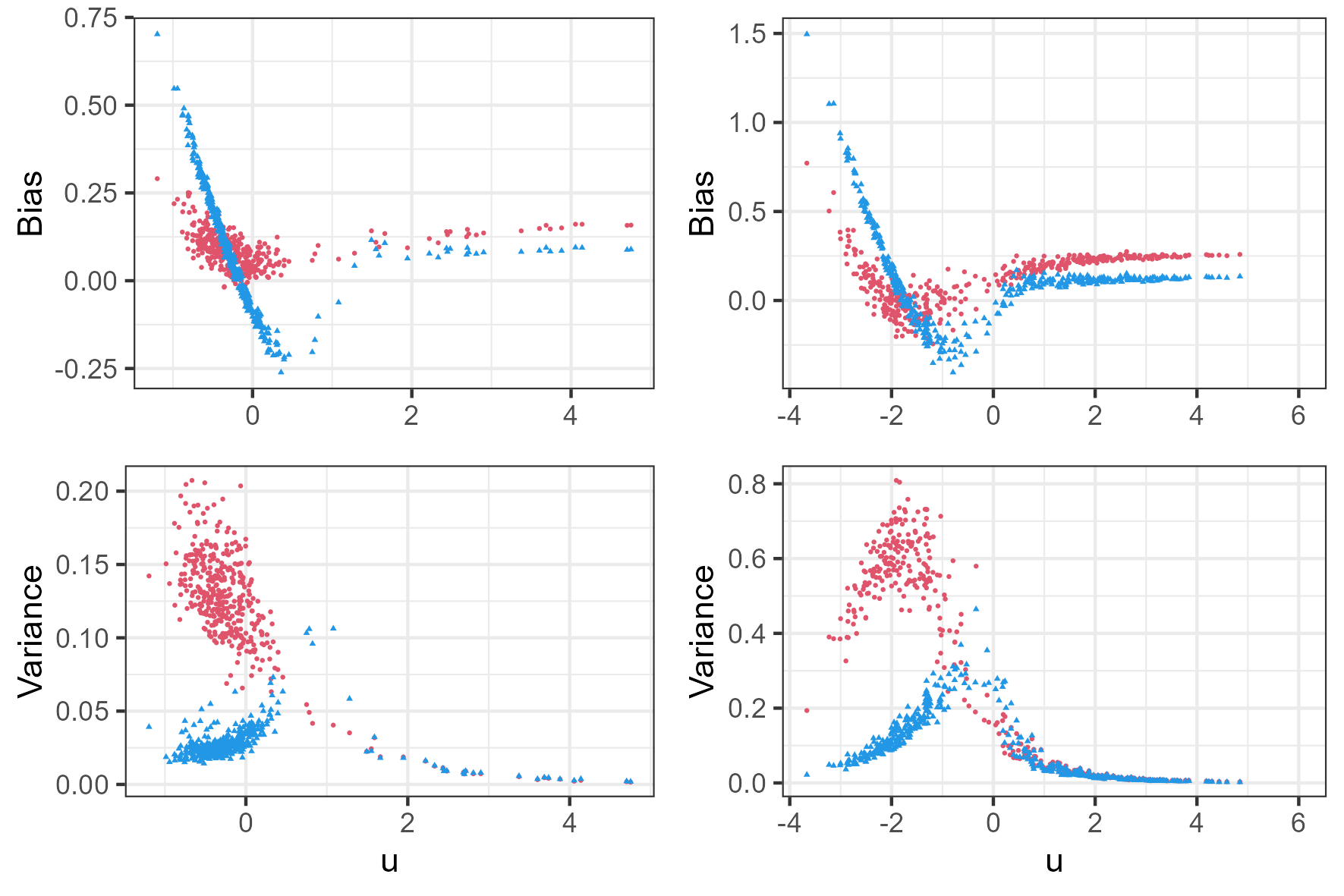}
\caption{
Bias and variance decomposition of the simulated CMSEP for the Poisson GLMM simulations. Red: Under the misspecified distribution. Blue: Under the correctly specified distribution. Left panel: True distribution is Distribution 1. Right panel: True distribution is Distribution 3.}
\end{figure}

\begin{figure}[h]
\centering
\includegraphics{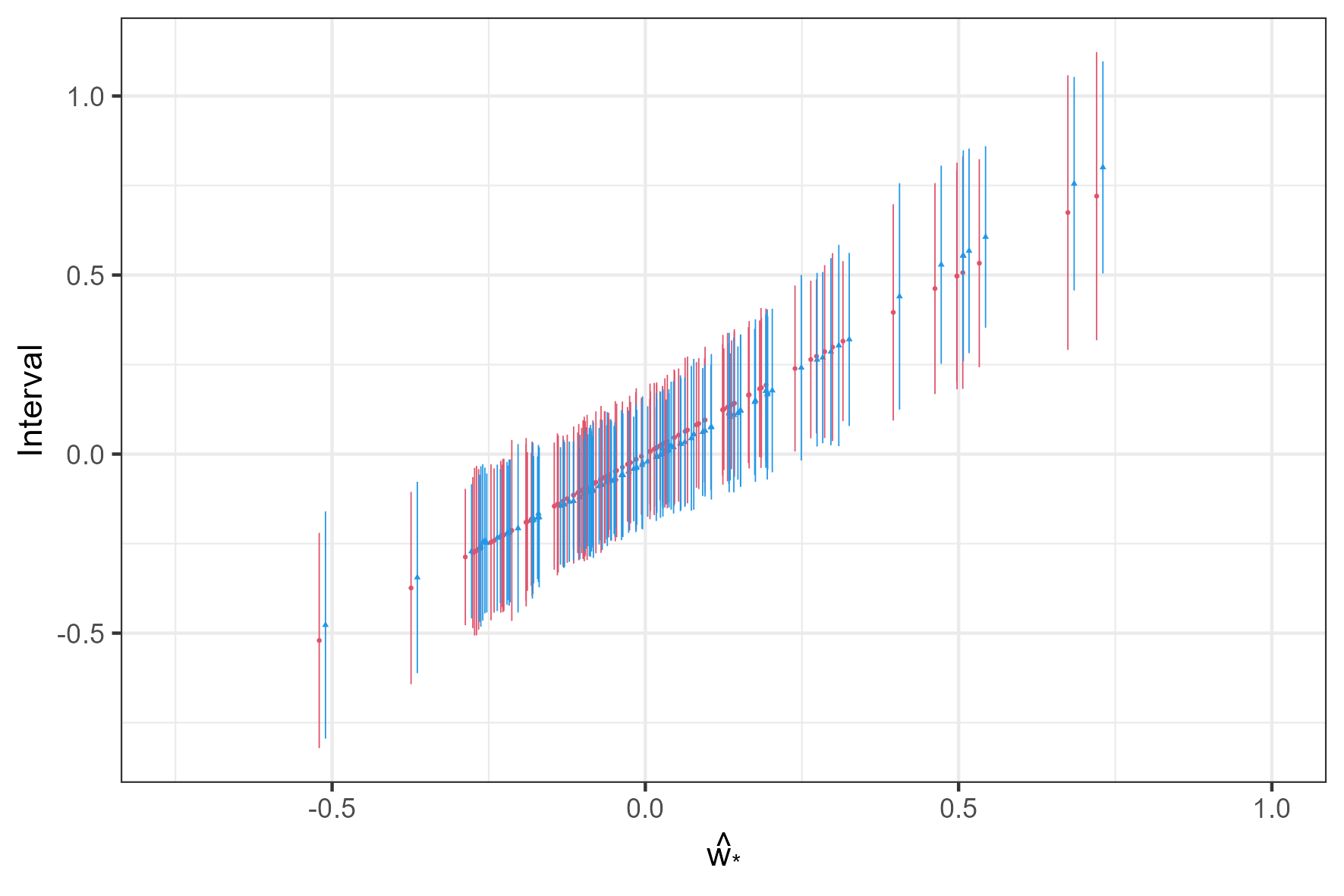}
\caption{Comparison of prediction intervals under LMMs fitted for the wages data assuming either the normal or mixture of normals random effects distribution. Red line: Intervals under the normal distribution. Blue line: Intervals under the mixture of normals distribution.}
\label{fig:wages3}
\end{figure}

\FloatBarrier

\subsection{Simulation study for LMMs}\label{app:sim_lmm}

We consider a linear mixed model with a single random intercept, $\{u_i, i = 1,\dots,m\}$. We used $q_f = 2$ fixed effects, and set the parameters of the linear mixed model to be $\betavec = (0, 1)^\top, x_{ij2} \sim Unif(0,1)$, and $\tau_0 = 1$.
We consider two mixture distributions (I) $p_0(u_i \mid \sigmavec_0) = 0.9 \N(-0.28, 0.28^2) + 0.1 \N(2.56, 1.42^2)$, and (II) $p_0(u_i \mid \sigmavec_0) = 0.5 \N(-1.77, 0.59^2) + 0.5 \N(1.77, 1.18^2)$.
To compare the effects of the cluster size $n_i$ and the number of clusters $m$, we vary $n_i$ over 5, 10, 20, and 40, and $m$ over 25, 50, and 100.

Table \ref{tbl:lmm_umsep} compares the simulated UMSEPs under the misspecified and true mixture distributions in different simulation settings. 
We can see that in most simulation settings, the values of $\text{UMSEP}_{\text{sim}}$ are larger under the misspecified normal distribution, although the difference is more evident when the cluster size $n_i$ is smaller and the number of clusters $m$ is larger. We also see that the difference between the simulated UMSEPs under the two distributions is larger when the true distribution is Distribution 1 than when it is Distribution 3.

\begin{table}[tb]
	\centering
	\caption{Simulated UMSEPs ($\text{UMSEP}_{\text{sim}}$) for the misspecified LMM/true LMM for two true mixture of normal random effects distributions.} 
	\label{tbl:lmm_umsep}
	\bgroup
	\begin{tabular}{ ccccc }
		\toprule[1.5pt]
		      & $n_i = 5$ & $n_i = 10$ & $n_i = 20$ & $n_i = 40$ \\
		\cmidrule{2-5}
            & \multicolumn{4}{c}{Random effects distribution I} \\
            $m = 25$ & 0.1778/0.1312 & 0.1386/0.1101 & 0.0793/0.0677 & 0.0644/0.0608 \\
            $m = 50$ & 0.1767/0.1070 & 0.1067/0.0724 & 0.0625/0.0513 & 0.0406/0.0388 \\
            $m = 100$ & 0.1768/0.1067 & 0.0997/0.0658 & 0.0585/0.0468 & 0.0323/0.0298 \\\\
            & \multicolumn{4}{c}{Random effects distribution II} \\
            $m = 25$ & 0.3611/0.3537 & 0.2176/0.2171 & 0.2094/0.2091 & 0.1668/0.1663 \\
            $m = 50$ & 0.2710/0.2550 & 0.1767/0.1682 & 0.1363/0.1349 & 0.1021/0.1016 \\
\            $m = 100$ & 0.2233/0.2017 & 0.1414/0.1355 & 0.0956/0.0939 & 0.0680/0.0674 \\
                        \bottomrule[1.5pt]
	\end{tabular}
	\egroup
\end{table}

\begin{figure}[t]
\centering
\includegraphics[width=1\textwidth]{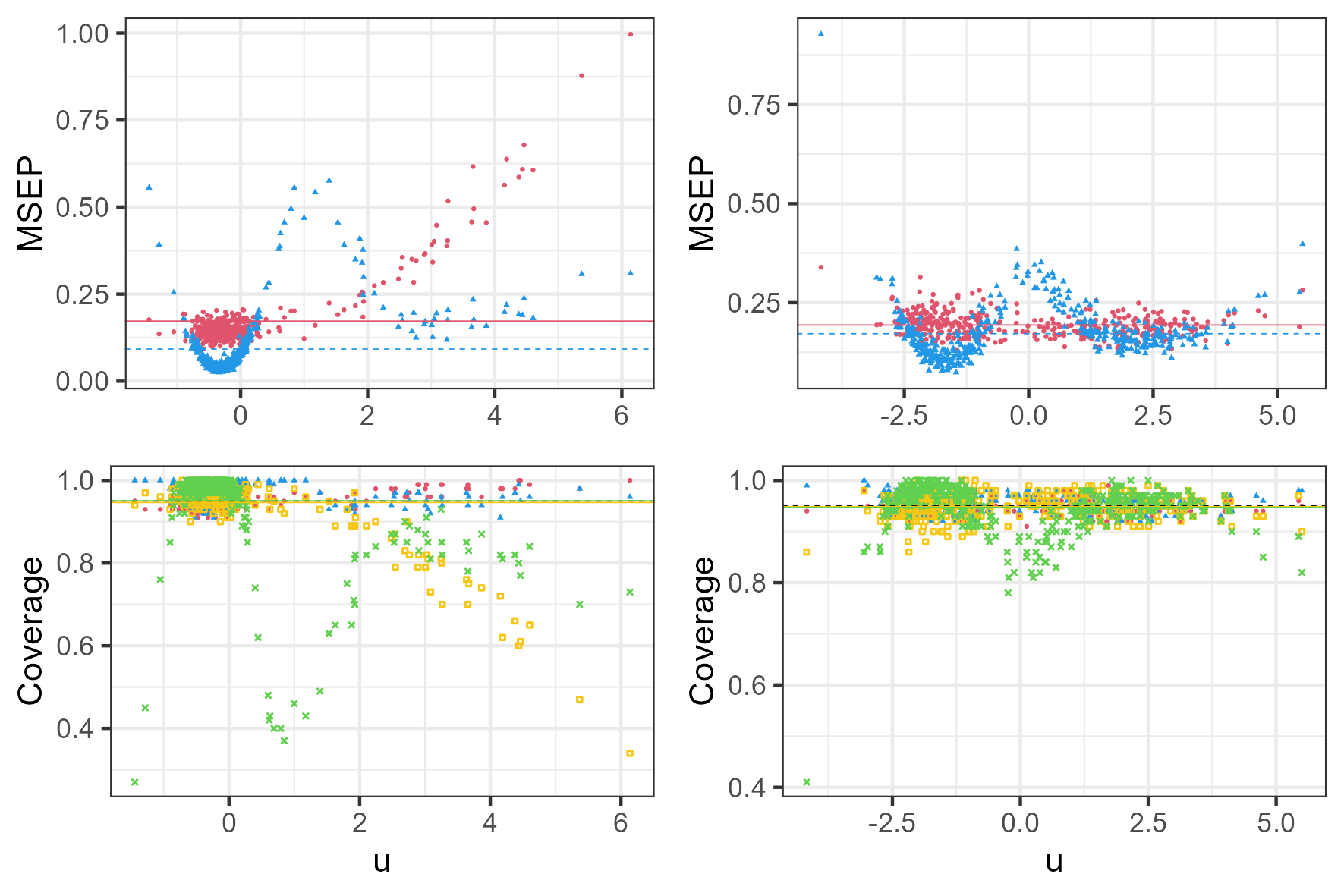}
\caption{Simulated CMSEPs ($\text{CMSEP}_{\text{sim}}$; top row) and empirical coverage probability of 95\% prediction intervals (bottom row) for the misspecified versus true LMMs, and for two true mixture of normal random effects distributions (left column is distribution I; right column in distribution II). 
In the top panel, the red dots and blue triangles correspond to the misspecified and true LMMs, while in the bottom panel the red dots and blue triangles correspond to intervals constructed using $\text{CMSEP}_{\text{sim}}$ under the misspecified and true LMMs, while the gold squares and green crosses correspond to intervals constructed using $\text{UMSEP}_{\text{sim}}$ under the misspecified and true LMMs.}
\label{fig:lmm_cmsep}
\end{figure}

Second, we investigate the conditional mean squared prediction error under the two random effects distributions. We fixed the cluster size $n_i = 5$ and the number of clusters $m = 400$.
Figure \ref{fig:lmm_cmsep} shows the OCMSEP on the random effects when the true random effects distributions are Distribution 1 and Distribution 3.
We see that under the misspecified model (the red dots), the value of $\text{CMSEP}_{\text{sim}}$ is lowest around zero (the assumed mean of the random effects distribution), and increases in the tail of the distribution, as detailed in Section \ref{sec:lmm_msep}.
Meanwhile, under the mixture distribution, for both Distribution 1 and Distribution 3, we see that there are two local minima for the observed CMSEP, one at the mean of each component of the corresponding true mixture distribution. This result is consistent with the discussion in Section \ref{sec:lmm_msep}. Around these regions, where the majority of the random effects are, the value of $\text{CMSEP}_{\text{sim}}$ is lower than under the misspecified GLMM.
The normality assumption can also impact the prediction in the tail of the random effects distribution, especially when the true distribution is skewed as occurs in Distribution 1. In that case, the value of $\text{CMSEP}_{\text{sim}}$ under the misspecified GLMM can be three to four times larger than under the true distribution, as shown in the left panel of Figure \ref{fig:lmm_cmsep}.

\end{document}